\documentclass[lettersize,journal]{IEEEtran}
\usepackage{amsmath,amsfonts}
\usepackage{algorithmic}
\usepackage{algorithm}
\usepackage{array}
\usepackage[caption=false,font=normalsize,labelfont=sf,textfont=sf]{subfig}
\usepackage{textcomp}
\usepackage{stfloats}
\usepackage{url}
\usepackage{verbatim}
\usepackage{graphicx}
\usepackage{cite}
\usepackage{color}
\usepackage{url}
\usepackage{epstopdf}
\usepackage{amsthm}
\usepackage{amsmath}
\usepackage{amsthm}
\usepackage{amssymb}
\usepackage{graphicx}
\usepackage{wrapfig}
\usepackage{algorithmic}
\usepackage{algorithm}
\usepackage{mathrsfs}
\usepackage{float}
\usepackage{mathtools}
\usepackage{booktabs}
\usepackage{caption}
\usepackage{multirow}
\usepackage{tabulary}
\usepackage{tablefootnote}
\usepackage{enumitem}

\usepackage{scalerel}
\usepackage{tikz}
\usetikzlibrary{svg.path}

\definecolor{orcidlogocol}{HTML}{A6CE39}
\tikzset{
  orcidlogo/.pic={
    \fill[orcidlogocol] svg{M256,128c0,70.7-57.3,128-128,128C57.3,256,0,198.7,0,128C0,57.3,57.3,0,128,0C198.7,0,256,57.3,256,128z};
    \fill[white] svg{M86.3,186.2H70.9V79.1h15.4v48.4V186.2z}
                 svg{M108.9,79.1h41.6c39.6,0,57,28.3,57,53.6c0,27.5-21.5,53.6-56.8,53.6h-41.8V79.1z M124.3,172.4h24.5c34.9,0,42.9-26.5,42.9-39.7c0-21.5-13.7-39.7-43.7-39.7h-23.7V172.4z}
                 svg{M88.7,56.8c0,5.5-4.5,10.1-10.1,10.1c-5.6,0-10.1-4.6-10.1-10.1c0-5.6,4.5-10.1,10.1-10.1C84.2,46.7,88.7,51.3,88.7,56.8z};
  }
}

\newcommand\orcidicon[1]{\href{https://orcid.org/#1}{\mbox{\scalerel*{
\begin{tikzpicture}[yscale=-1,transform shape]
\pic{orcidlogo};
\end{tikzpicture}
}{|}}}}

\usepackage{cite}
\usepackage{amsmath,amssymb,amsfonts}
\usepackage{graphicx}
\usepackage{textcomp}
\usepackage{xcolor}
\usepackage{hyperref}
\setlist[itemize]{leftmargin=*}
\usepackage{enumitem}

\begin{document}

%\title{A Survey on AI/ML-Driven Intrusion and Misbehavior Detection Systems in Networked Autonomous Vehicles: Techniques, Challenges and Opportunities}
\title{A Survey on AI/ML-Driven Intrusion and Misbehavior Detection in Networked Autonomous Systems: Techniques, Challenges and Opportunities}

\author{Opeyemi Ajibuwa\textsuperscript{\orcidicon{0009-0007-5956-0925}}, Bechir Hamdaoui\textsuperscript{\orcidicon{0000-0002-6085-4505}},~\IEEEmembership{Senior Member, IEEE}, Attila A. Yavuz,~\IEEEmembership{Senior Member, IEEE}
        % <-this % stops a space
\thanks{This work was supported in part by US National Science Foundation (NSF) under awards No. 1923884 and No. 2003273. \em{(Corresponding author: Opeyemi Ajibuwa.)}}
\thanks{Opeyemi Ajibuwa and Bechir Hamdaoui are with the School of Electrical Engineering and Computer Science, Oregon State University, Corvallis, OR 97333, USA (e-mail: ajibuwao@oregonstate.edu; hamdaoui@oregonstate.edu).}% <-this % stops a space
\thanks{Attila A. Yavuz is with Department of Computer Science and Engineering, University of South Florida, Tampa, FL 33620, USA (e-mail: attilaayavuz@usf.edu).}
}

\markboth{}
{Shell \MakeLowercase{\textit{et al.}}: A Sample Article Using IEEEtran.cls for IEEE Journals}

% \IEEEpubid{0000--0000/00\$00.00~\copyright~2023 IEEE}

\maketitle

\begin{abstract}
AI/ML-based intrusion detection systems (IDSs) and misbehavior detection systems (MDSs) have shown great potential in identifying anomalies in the network traffic of networked autonomous systems. Despite the vast research efforts, practical deployments of such systems in the real world have been limited. Although the safety-critical nature of autonomous systems and the vulnerability of learning-based techniques to adversarial attacks are among the potential reasons, the lack of objective evaluation and feasibility assessment metrics is one key reason behind the limited adoption of these systems in practical settings.  This survey aims to address the aforementioned limitation by presenting an in-depth analysis of AI/ML-based IDSs/MDSs and establishing baseline metrics relevant to networked autonomous systems. Furthermore, this work thoroughly surveys recent studies in this domain, highlighting the evaluation metrics and gaps in the current literature. It also presents key findings derived from our analysis of the surveyed papers and proposes guidelines for providing AI/ML-based IDS/MDS solution approaches suitable for vehicular network applications. Our work provides researchers and practitioners with the needed tools to evaluate the feasibility of AI/ML-based IDS/MDS techniques in real-world settings, with the aim of facilitating the practical adoption of such techniques in emerging autonomous vehicular systems.

%Researchers are more focused on presenting theoretical proposals than evaluating the feasibility of their solutions in real-world scenarios.

\end{abstract}


\section{Introduction}
%\IEEEPARstart{B}{reakthroughs} in sensing, computing, and communication technologies have transformed cyber-physical systems \cite{rajkumar2010cyber, rajkumar2012cyber, serpanos2018cyber}, including the subdomains of Autonomous Vehicles (AVs) and Unmanned Aerial Vehicles (UAVs) \cite{song2017security}. The reasonable cost of computing and sensing and support of 5G wireless services have led to tremendous growth and innovation in these systems. 
%
\IEEEPARstart{B}{reakthroughs} in sensing, computing, and communication technologies have transformed the realm of cyber-physical systems (CPS) \cite{rajkumar2010cyber, rajkumar2012cyber, serpanos2018cyber}. Autonomous Vehicles (AVs) and Unmanned Aerial Vehicles (UAVs) \cite{song2017security} are two subdomains of such systems that have seen tremendous growth and innovation due to the reasonable computing/sensing costs and the support of 5G network services.  With the increasing number of devices and expanding use cases, critical data must be shared between different system components to enable proper functioning and safety. 
%
%Additionally, there is a growing demand for vehicle-to-everything (V2X) communication and its variants to ensure safety in interactions between vehicles and infrastructure \cite{sun2015cluster, siegel2017survey}, and Flying Ad Hoc Networks (FANETs) and their variants for UAV swarms/networks \cite{bekmezci2013flying, quaritsch2010networked}. The proliferation of these networks has expanded connectivity, enabling numerous applications to operate effectively. 
%
Additionally, there has been a growing demand for vehicles to interact with one another and with other infrastructures, for efficiency and safety reasons \cite{sun2015cluster}. As a result, concepts such as Vehicle-to-Everything (V2X) communications and Flying Ad-Hoc Networks (FANETs) and their variants have emerged \cite{siegel2017survey,bekmezci2013flying, quaritsch2010networked}. 

The proliferation of these emerging networks and services has expanded connectivity, enabling numerous applications to operate effectively. However, this has also created new vulnerabilities for attackers to exploit. In an effort to overcome these challenges, recent years have witnessed the development of many intrusion detection systems (IDS) \cite{raya2007eviction, kumar2014collaborative, sedjelmaci2013efficient, mitchell2013adaptive} and misbehavior detection systems (MDS) \cite{grover2011machine, van2013misbehavior, ruj2011data}, with a focus on the use of Artificial Intelligence/Machine Learning (AI/ML) tools, to identify and prevent attacks on these safety-critical systems. 
There has been a significant increase in the proposed AI/ML-based IDS/MDS solutions, due to their high accuracy in detecting complex patterns in large datasets. While these systems have demonstrated admirable performance in research studies, their full potential has yet to be realized in real-world scenarios and settings \cite{sommer2010outside}.

Given the success of AI/ML approaches in other domains \cite{casas2020two, sarker2021machine} and the promise of AI/ML-based IDS/MDS in networked autonomous systems, previous works \cite{ashraf2020novel, xie2021threat, uprety2021privacy} might not have shown their real-world viability at a full extend. While conducting experiments to test the effectiveness of these systems is prohibitively expensive, little effort has been made in the literature to assess their feasibility for real-world applications. Given the highly dynamic and mobile nature of autonomous system environments, there is a critical need for high-performing security solutions that are also feasible and practical. This work aims to fill this gap by providing a comprehensive examination of previously proposed IDS and MDS solutions for AV and UAV networks, enabling researchers and industry practitioners to make informed feasibility assessments of these proposed frameworks.

\IEEEpubidadjcol
This paper establishes baseline requirements to facilitate the feasibility assessment of IDS and MDS techniques in AV and UAV networks. By reviewing the literature, key metrics are identified to validate proposals for IDS/MDS in real-world AV/UAV networks. While prior surveys in this domain have attempted to highlight the limitations of AI/ML-based IDS/MDS proposals, this paper broadens the scope to identify challenges that limit these intrusion and misbehavior detection systems. Additionally, potential directions for addressing these challenges in networked autonomous systems are presented.

\renewcommand{\arraystretch}{1.2}
\begin{table*}[htb]
\caption{\small Summary of AI/ML-based techniques, scope, and comparison of related surveys} 
\begin{tabulary}{\textwidth}{LCCCL}
    \toprule \toprule
    RELATED SURVEYS\!\! & AREA & \!\!\!\!YEAR & SYSTEM & DIFFERENCE FROM OUR SURVEY \\ 
    \hline
    Deep AI/ML-based Anomaly Detection in Cyber-physical Systems: Progress and Opportunities \cite{luo2021deep} & Anomaly detection & 2021 &
    CPS & Focuses on DL-based CPS security. Our work broadens the scope to encompass other types of learning and presents some new directions and novel work in those domains. \\ 
    \hline
    Recent Advances in Machine-Learning Driven Intrusion Detection in Transportation: Survey \cite{bangui2021recent} & Intrusion detection & 2021 & VANETs and UAVs & Brief review of IDS applications in UAV networks. Ours went into greater depth surveying AI/ML-based IDS/MDS approaches in both AV and UAV networks. \\ 
    \hline
    A Survey on Machine Learning-based Misbehavior Detection Systems for 5G and Beyond Vehicular Networks \cite{boualouache2022survey} & Misbehavior detection &
    2022 & 5G and VANETs &
    Focuses on IDS techniques only. Our work covers both IDS and MDS approaches, as well as highlights novel challenges these systems face. \\ 
    \hline
    Machine Learning for Security in Vehicular Networks: A Comprehensive Survey \cite{talpur2021machine} & Security, Trust and Privacy & 2021 & VANETs and its variants &
    Our work focuses on AI/ML-based IDS/MDS techniques for AV and UAV networks.
    \\ \hline
    A Survey on Machine Learning-based Misbehavior Detection Systems for 5G and Beyond Vehicular Networks \cite{boualouache2023survey} & Misbehavior detection & 2023 &  5G and VANETs & Reviews MDS techniques in VANETs and 5G networks only. Our work more broadly reviews IDS and MDS applications in VANETs and UAVs. \\ 
    \hline
    This work & Intrusion and Misbehavior detection & -- &  UV and UAV networks & --  \\
    \bottomrule \bottomrule
\label{tab:summary}
\end{tabulary}
\end{table*}

\subsection{State-of-the-Art Surveys}
Previous surveys on the AL/ML-based IDS/MDS literature for vehicular and related systems mostly fail to provide a thorough assessment of the reviewed techniques vis-a-vis of their feasibility and adoption in real-world settings (see Table \ref{tab:summary}). 
%Additionally, most of the IDS/MDS research has focused on improving detection accuracy rather than optimizing these systems for practical deployments. 
%
Luo et al. \cite{luo2021deep} reviewed AI/ML-based anomaly detection approaches for cyber-physical systems, categorizing works by threat models, detection strategies, implementation, and evaluation metrics. They discussed neural network models for building deep anomaly detection and limitations and weaknesses of existing methods but did not evaluate the feasibility of these approaches in real-time applications. Bangui et al. \cite{bangui2021recent} surveyed ML-based techniques for IDS in VANETs and UAV networks and discussed some known challenges, but did not address key challenges (e.g., concept drift, adversarial attacks), nor did they provide a feasibility assessment.

Boualouache et al. \cite{boualouache2022survey} provided a detailed survey on ML-based MDS for V2X communications, discussing misbehavior detectors from security and ML perspectives and offering recommendations for developing, validating, and deploying these ML-based MDS approaches. While highlighting open research and standardization issues, they did not evaluate the feasibility of these works in real-world settings.

Talpur et al. \cite{talpur2021machine} systematically reviewed ML-based techniques for addressing security issues in vehicular networks, providing a taxonomy of attacks and discussing the associated security challenges and requirements. However, the feasibility of the existing works was not discussed.

Boualouache et al. \cite{boualouache2023survey} conducted a survey of ML-based MDS approaches for 5G and beyond vehicular networks. In their review, the authors offered valuable recommendations for developing, validating, and deploying ML-based MDS solutions, while also emphasizing standardization issues and potential research directions. However, it should be noted that their survey was limited to MDSs in 5G and VANETs, and did not cover other networked autonomous systems.

%Given the gaps in previous surveys, current AI/ML-based IDS/MDS coverage may be inadequate, especially regarding the proposed techniques' practical feasibility in safety-critical networked systems (e.g., AV/UAV networks). Currently, there are no established baselines against which proposed solutions could be compared. Our work aims to address this gap by reviewing recent works and highlighting their current limitations based on our findings. Lastly, we provide new directions from other domains to facilitate the adoption of AI/ML-based IDS/MDS solutions in real-world scenarios. Our goal is to establish baseline assessment metrics that encourage future research innovations in this field.

\subsection{Our Contributions}
While previous surveys have attempted to review the relevant literature in this area, none of them considered the feasibility of the proposed techniques in these safety-critical networked autonomous systems (see Table \ref{tab:summary}). 
%Additionally, most of the IDS/MDS research has focused on improving detection accuracy rather than optimizing these systems for practical deployments. 
Additionally, most IDS/MDS research has focused on improving detection accuracy with little consideration for real-world deployment efficiency and optimality. This work fills this gap by providing baseline standards for evaluating the feasibility of AI/ML-based IDS and MDS for AV/VANET and UAV networks. We review recent works, highlight their limitations, and present new directions for tackling the identified challenges to encourage future research innovations. Our goal is to set baseline assessment metrics for AI/ML-based IDS/MDS solutions to facilitate their adoption in real-world scenarios. The following summaries of our contributions:

\begin{itemize}[leftmargin=2em]
\item  We provide a comprehensive review of the taxonomy of AI/ML-based intrusion and misbehavior detection systems in AV/UAV networks. This includes the examination of their architectures, threat models, and other considerations to establish baseline feasibility metrics for assessing IDS/MDS in networked autonomous systems.

\item To the best of our knowledge, we are the first to formally define baseline assessment metrics for critically evaluating the feasibility of IDS/MDS solutions for networked autonomous systems in the literature.

\item We conduct a deep analysis of existing AI/ML-based IDS and MDS solutions, based on the feasibility metrics. Additionally, we identify exciting research in the general domain of AI/ML and provide suggestions for adopting these solutions to tackle the identified challenges.
\end{itemize}

\subsection{Organization and Acronyms}
\begin{figure*}
\centerline{\includegraphics[width=.9\linewidth]{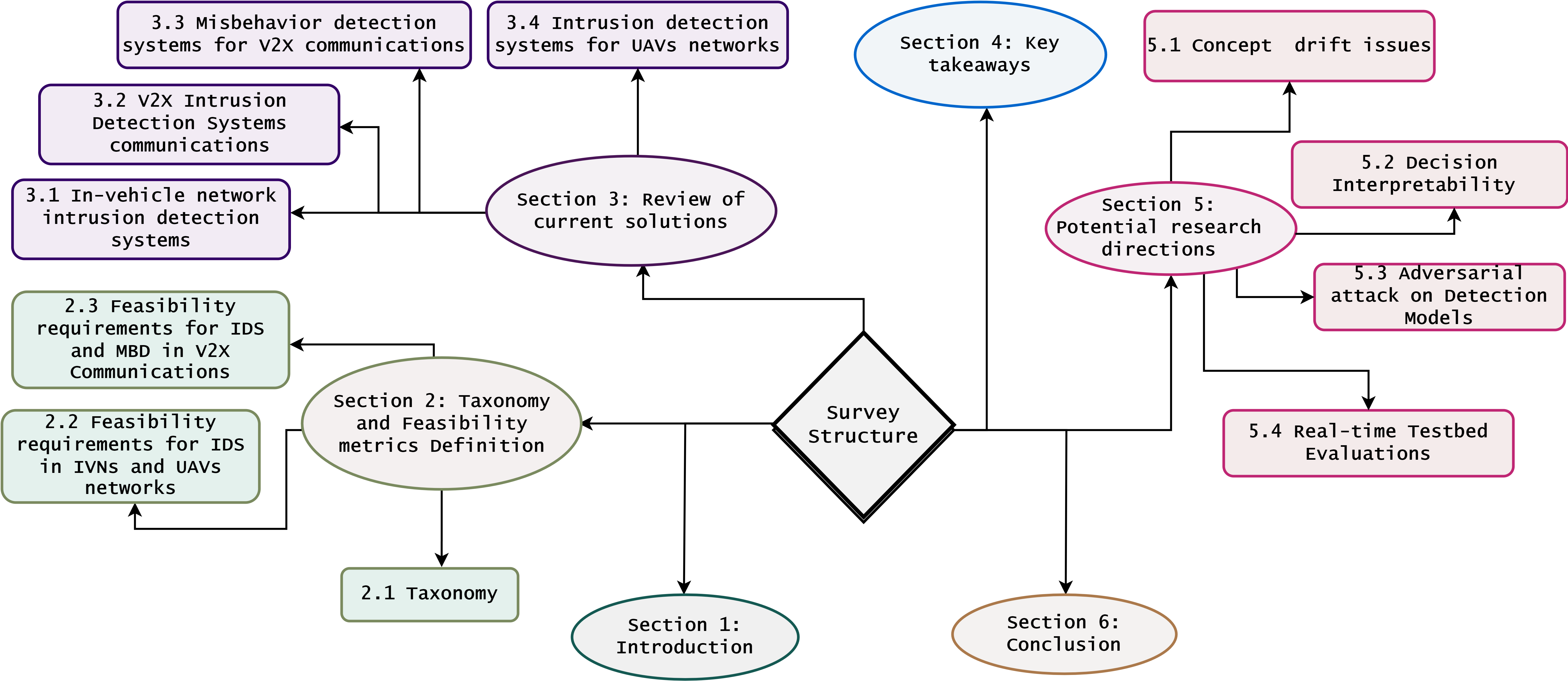}}
\caption{The organization and structure of the survey}
\label{figstruct}
\end{figure*}

The rest of the survey is structured as follows, as shown in \autoref{figstruct}. The taxonomy, architecture, and baseline metrics are presented in \autoref{sec:background}. \autoref{sec:evaluation} analyzes each of the current works in terms of the established metrics for real-world applications. A detailed review of the limitations of the current works is provided in \autoref{sec:inference}. \autoref{sec:Advances} presents new ideas and directions to address some of the challenges with current AI/ML-based IDS/MDS solutions. Finally, the paper concludes in \autoref{sec:conclusion}. A
summary of the acronyms used is provided in Table \ref{tab:acronym}.

\begin{table}
\caption{\small Acronyms used throughout the paper} 

\begin{center}
\begin{tabular}{ll}
\toprule
    \textit{ML} & Machine Learning                 \\
    \textit{DL}  & Deep Learning                    \\
    \textit{VANETs} & Vehicular Ad Hoc Networks        \\
    \textit{FANETs} & Flying Ad Hoc Networks           \\
    \textit{CAN} & Controller Area Networks         \\
    \textit{V2X} & Vehicle-to-Everything            \\
    \textit{V2V} &  Vehicle-to-Vehicle               \\
    \textit{V2I} & Vehicle-to-Infrastructure        \\
    \textit{V2R} & Vehicle-to-Road                  \\
    \textit{V2G} & Vehicle-to-Grid                  \\
    \textit{V2C} & Vehicle-to-Cloud                 \\
    \textit{AVs} & Autonomous Vehicles               \\
    \textit{UAVs} & Unmanned Aerial Vehicles            \\
    \textit{BSM} & Basic Safety Messages            \\
    \textit{CAMs} & Cooperative Awareness Messages   \\
    \textit{ECUs} & Electronic Control Units         \\
    \textit{MiTM} & Man-in-the-Middle                \\
    \textit{MBD} &  Misbehavior Detection            \\
    \textit{MDS} &  Misbehavior Detection System     \\
    \textit{OBD} &  Onboard Diagnostics Units        \\
    \textit{DLC} &  Diagnostics Link Connector       \\
    \textit{FNR} &  False Negative Rate              \\
    \textit{FPR} &  False Positive Rate              \\
    \textit{LOS} &  Line-of-Sight                    \\
    \textit{IoT} &  Internet of Things               \\
    \textit{RSU} &  Roadside Units                   \\
    \textit{MEC} &  Multi-Access Edge Cloud          \\
    \textit{OBUs} & On-board Units                   \\
    \textit{SDNs} & Software Defined Networks        \\
    \textit{OEMs} & Original Equipment Manufacturers \\
    \textit{IDS} & Intrusion Detection System       \\
    \textit{IVN} & In-Vehicle Network               \\
    % \textit{IoD} & Internet of Drones               \\
\bottomrule
\label{tab:acronym}
\end{tabular}
\end{center}
\end{table}

\section{Background and Feasibility Metrics of IDS/MDS Solution Approaches}
\label{sec:background}
We first explain our taxonomy for AI/ML-based IDS/MDS techniques in VANET/AV networks and UAV networks. We then present an overview of the architectures and requirements for assessing the feasibility of the proposed solutions in each of the studied domains: in-vehicle networks (IVN), V2X communications, and UAV networks. Finally, we establish the metrics for evaluating the viability of existing IDS and MDS techniques based on these requirements.

\subsection{Taxonomy of AI/ML-based IDS and MDS Approaches}
\begin{figure*}[ht]
\centerline{\includegraphics[width=1.0\linewidth]{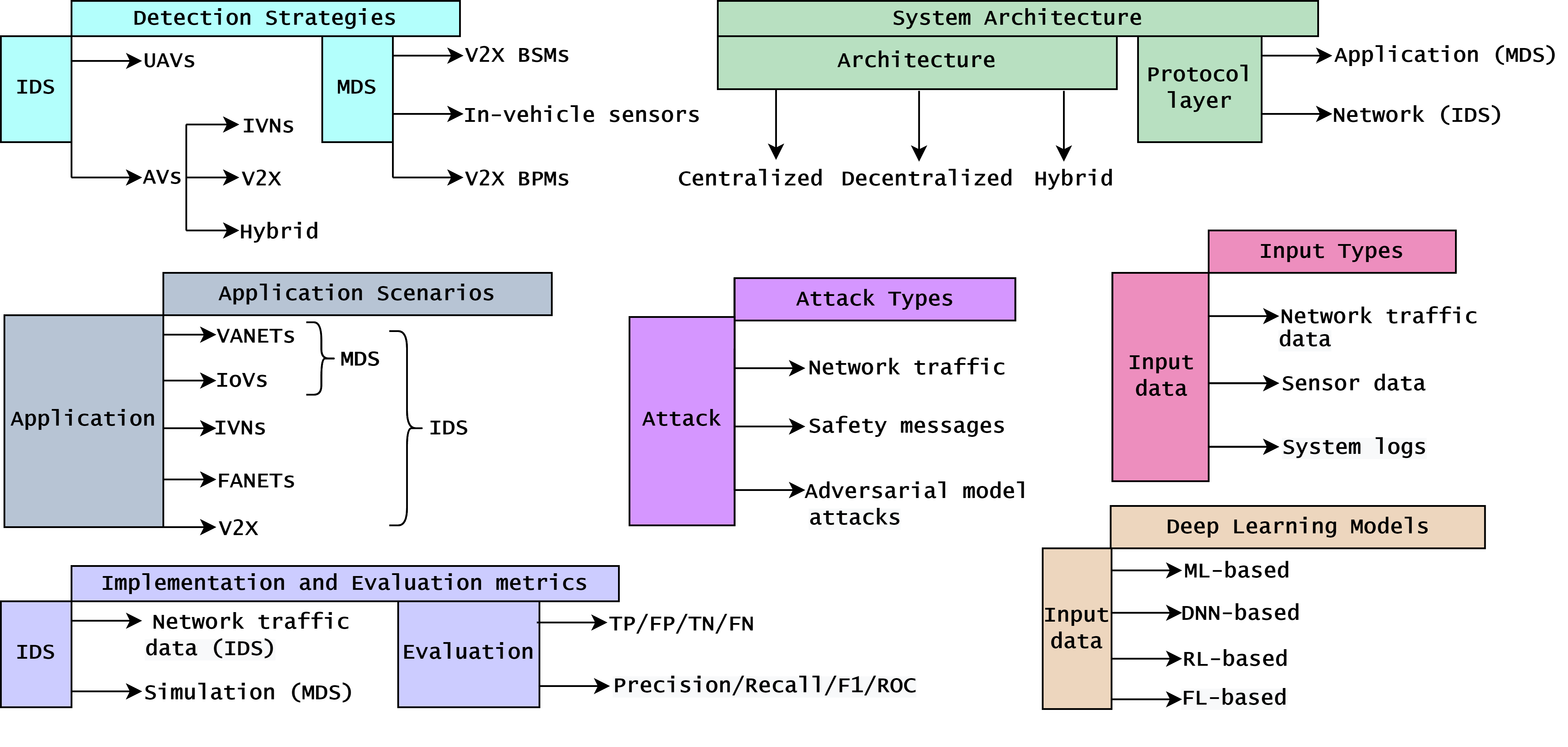}}
\caption{Taxonomy of this survey.}
\label{fig1}
\end{figure*}

We depict our taxonomy in \autoref{fig1}, which consists of (1) detection strategies, (2) systems architectures, (3) application scenarios, (4) attack and input types, (5) machine learning models, and (6) implementation and evaluation methods.

\subsubsection{Detection Strategies} 
Based on our scope, we mainly examine intrusion detection and misbehavior detection. While both VANETs and UAV networks make use of IDSs, MDSs are mainly utilized in vehicles \cite{choudhary2018intrusion}. 
It could be because insider attacks in UAV swarms are more difficult to execute in comparison to vehicular networks. The following is a further breakdown of these systems:

%The reason for this could be attributed to the fact that insider attacks in UAV swarms are more difficult to execute in comparison to vehicular networks. The following is a further breakdown of these systems:

\begin{itemize}[leftmargin=2em]
    \item {\em IDS for In-vehicle Networks (IVNs):} Modern cars' complexity and connectivity pose significant security threats to vehicular networks \cite{wu2019survey}. Additionally, certain applications may lack broadcast authentication and messaging \cite{al2019intrusion}, which is why an IDS must be in place to monitor and detect any abnormalities in the CAN-BUS network.
    \item {\em IDS for V2X Communications:} Exposure to multiple connection points makes vehicles vulnerable to various attacks that could threaten lives and other essential infrastructures. To prevent these attacks, IDSs integrate cryptographic security techniques into various V2X communications (e.g., V2V, V2I, V2R, etc.) \cite{yang2021mth}.
    
    \item {\em IDS for UAVs and Hybrid Networks:} Connected UAVs are also vulnerable to attacks, some of which can be detected via IDS integrated into UAV's communication network \cite{abu2022high}. Hybrid IDSs are designed to protect vehicular networks from both internal and external attacks.
    
    \item {\em MDS for V2X Basic Safety Messages:} Misbehavior detection for V2X communications involves monitoring the data semantics \cite{kamel2020simulation} of transmitted Basic Safety Messages (BSMs) to detect attempts that aim to sabotage the network by sending bogus messages that could mislead other vehicles.
    
    \item {\em MDS for In-vehicle Sensory Data:} Detecting anomalies in vehicles caused by incorrect sensor data readings is critical for overall system safety. Although faulty sensor data may not be planned as an attack, detecting these anomalies early on is essential to prevent them from being communicated to other vehicles and misinterpreted as an act of attack.
    
    \item {\em MDS for V2X Basic Perception Messages:} Semantic validity of Collective Perception Messages (CPMs) is essential to prevent attacks on V2X systems \cite{liu2021miso}. Trusting CPMs is risky, and as such, IDSs have been proposed to verify their semantic validity.
\end{itemize}

\subsubsection{System Architectures} IDS can be deployed in three architectures: centralized, decentralized, and hybrid. A centralized architecture involves aggregating data at a central node for attack detection and model training. In a decentralized architecture, training, and detection are distributed across local nodes in the network. In a hybrid architecture, models are locally trained on each node and centrally aggregated to provide network-wide attack detection capabilities. In VANETs, MDS operates as an application because misbehavior occurs at the application layer of the communication protocol stack. On the contrary, IDS addresses traffic intrusion at the network layer.

\subsubsection{Application Scenarios} With regards to applications, IDSs are utilized in both AV and UAV networks, while MDSs are limited to specific AV network cases, as outlined next: 

\begin{itemize}[leftmargin=2em]
    \item {\em VANETs and Internet of Vehicles (IoV):} VANETs are limited to vehicle-to-vehicle communication, it is an ad-hoc network where vehicles are connected and exchange data with each other \cite{verma2021evolution}. In contrast, IoV encompasses a more extensive network involving the real-time exchange of information among vehicles, sensors, roads, and personal devices \cite{gasmi2019vehicular}.  These information exchanges should ideally be authenticated by respecting the real-time (i.e., delay-aware) requirements of IoVs~\cite{7953565}. 
    
    \item {\em Autonomous Vehicles (AVs):} AVs, with their vast network of ECUs and sensors, are susceptible to both internal and external threats. In contrast to conventional vehicles, AVs must communicate with external networks, such as other vehicles and infrastructure, which could potentially serve as a channel for attackers and hackers \cite{kumar2018brief}.
    
    \item {\em FANETs and Internet of Drones (IoD):} FANETs are a type of network comprising a group of small unmanned aerial vehicles (UAVs) connected in an ad hoc manner and integrated into a team to achieve high-level goals \cite{khan2017flying}. On the other hand, IoD~\cite{8599784} is a network architecture that allows users to communicate and control UAVs through the internet.
\end{itemize}

\subsubsection{Attack and Input Types} VANETs and UAV networks are vulnerable to various types of attacks targeting different aspects. Attackers may carry out a man-in-the-middle attack or a denial-of-service attack on the network traffic. Application-layer attacks can also occur, such as on the safety messages of the network, or on the intrusion or misbehavior detection models. MDS are fed sensor data or BSMs, while IDS either parses traffic logs or network traffic for detection.

\subsubsection{Machine Learning Models} For IDS/MDS in VANETs and UAV networks, the most commonly used ML models are: deep neural networks (DNN) \cite{hossain2020effective, xie2021threat, ashraf2020novel}, reinforcement learning (RL) \cite{bouhamed2021lightweight, xiong2020reinforcement, sedar2021reinforcement} and federated learning (FL) \cite{uprety2021privacy}.

\subsubsection{Implementation and Evaluation Methods} ML models for IDS are trained and tested using time-series network traffic data, while MDS are often trained and evaluated using BSMs or sensory data obtained from simulation tools like VEINS \cite{sommer2019veins} and CARLA \cite{dosovitskiy2017carla}. Accuracy, precision, recall, and Receiver Operating Characteristics (ROC) are the most commonly used metrics for evaluating such systems \cite{yang2021lightweight, ercan2021misbehavior, sharma2020machine, gyawali2020machine}.

\subsection{Feasibility Requirements for Intrusion Detection Systems in IVNs and UAV Networks}
To assess the feasibility of IDS in IVNs and UAV networks, this section raises a set of key questions, the answers to which  will help derive metrics for characterizing the effectiveness of IDS in these contexts. These questions are presented next:

\subsubsection{What are the real threats to IVNs} 
Several solutions have been proposed in the literature based on different attacker models \cite{javed2021canintelliids, song2020vehicle, ashraf2020novel}, but some of these attacks lack evidence or have flawed assumptions \cite{sharma2021detailed}, making it necessary to focus on attacks that pose a real threat to IVN and for which IDS solutions are applicable. Misbehavior in IVN often arises from erroneous sensor readings \cite{gyawali2020machine}, which are outside the scope of this study. However, it has been shown that an IVN can be compromised either by plugging into the OBD-II port or remotely accessing an ECU via WiFi, Bluetooth, or telematics services \cite{koscher2010experimental}. The attacks that have been demonstrated to pose serious threats to IVNs are briefly described next.

\begin{itemize}[leftmargin=2em]
    \item {\em Denial-of-Service (DoS) attacks:} The attacker spams the IVN by sending messages with a high priority and at a fast rate, delaying or completely denying access to other legitimate nodes that need to send messages on the network. The works in \cite{hossain2020effective, lin2020evolutionary, hanselmann2020canet, tariq2020cantransfer} have demonstrated the potency of this attack.
    \item {\em Spoofing attacks:} In this attack, an attacker can create and inject spoofed messages into the IVN to impersonate a victim's ECU. This attack is easily realized by checking the DBC file for a particular car or by performing reverse engineering on the network bus \cite{zhang2022hybrid, lin2020evolutionary}.
    \item {\em Replay attacks:} The replay attack injects previously captured traffic into the network and masks it as regular data, slowing real network traffic. Although the replayed traffic represents a previously legitimate traffic sequence, the delay associated with its later insertion disrupts the network's real-time sequence of events. Previous tests showed that this can be easily carried out in most automobiles \cite{zhang2022hybrid, lokman2019intrusion, tariq2020cantransfer, bi2022intrusion}.
    
    \item {\em Fuzzing attacks:} This attack randomly injects compromised CAN ID, DLC, and data fields, resembling legitimate traffic, to manipulate the IVN. The disruptions caused by this attack have been reported to result in abnormalities such as abnormal engine noise \cite{lee2017otids}, abnormal increase in power \cite{lee2017otids}, gear switching \cite{bi2022intrusion}, erratic change of signal lights \cite{hossain2020long}, etc.
    
    \item {\em Drop attacks:} The drop attack manifests itself by a compromised ECU pausing or terminating packet transmission. This implies that messages from the compromised ID do not appear in network traffic for some time after it has been hijacked. This attack has been demonstrated in tests by \cite{miller2015remote, foster2015fast, hanselmann2020canet}.
    
    \item {\em Malware:} Malicious software could be introduced into the IVN through various means, such as infected USB drives or through wireless connections. Malware could be designed to steal sensitive information, disrupt vehicle operations, or compromise the driver's safety ]\cite{kang2016intrusion, liu2017vehicle, derhab2021histogram}.
    
    \item {\em Eavesdropping:} IVNs lacking proper encryption are vulnerable to eavesdropping, where attackers could intercept and decode sensitive information~\cite{glas2012signal}, such as GPS coordinates or vehicle speed, to gain insights into the driver's behavior and/or to launch malicious attacks \cite{ning2019attacker, khatri2021security}. There are encryption methods specially tailored for IVNs to mitigate some of these attacks, especially considering CAN bus protection~\cite{merchan2015system}.  
    
    \item {\em Social engineering:} Attackers could use social engineering techniques to trick drivers or passengers into giving them access to the IVN. For example, attackers could pose as maintenance personnel or IT support personnel to gain access to the vehicle \cite{luo2021research, kleberger2011security, costantino2018candy}.
    \item {\em Other attack variants:} Some other forms of attacks reported on the IVN include ulterior fuzzy attacks \cite{bi2022intrusion}, plateau attack \cite{hanselmann2020canet}, continuous change attack \cite{hanselmann2020canet} and malfunction attack \cite{han2021ppm}.
\end{itemize}

%\begin{figure*}[htb]
%\centering
%\includegraphics[width=1.0\linewidth]{figures/ivn.png}
%\caption{(a) Overview of ECU, Sensor and Actuator as illustrated in \cite{wu2019survey}. (b) Schematic %diagram of a typical in-vehicle network architecture of a modern automobile as illustrated in %\cite{sagstetter2013security}.}
%\label{fig2}
%\end{figure*}

\subsubsection{Are current vehicle architectures fit to meet the requirements for IVN IDS}
This section briefly examines specific issues that affect the practical fitness of IVN IDS solutions.

\begin{itemize}[leftmargin=2em]
    \item {\em Lack of industry standards for IDS deployment in IVN:} The lack of detailed specifications for IDS requirements in modern vehicle designs poses a key challenge to IVN IDS deployments. Although the ISO 21434 standard \cite{dantas2020security} provides general cybersecurity design guidelines for IVN, it doesn't provide the necessary details to develop and deploy IDS for automotive use. As of now, there are no industry standards for defining the infrastructure requirements for deploying an IDS in IVNs.
    
    \item {\em Integration issues:} Contemporary IDS proposals do not account for modern vehicle designs, which result in integration problems for many of the IVN IDS solutions described in the literature. Given the variability of network resources and performance requirements among the ECUs in an IVN, it is essential to consider the placement of ECUs and their real-time requirements while developing an IDS for the network.
    
    \item {\em ECUs architecture and latency demands:} The architecture of ECUs in vehicles typically is divided into different domains based on their function and latency requirements. High-end vehicles have at least 70 working ECUs, spread throughout the vehicle and interconnected at the gateway. Recommendations for IDS deployment to protect ECUs vary, with some suggesting an IDS for each ECU and others for the bus gateway. Real-time requirements of different domains that make up an IVN must be considered to decide which architecture is best. Safety-critical components, such as powertrain and chassis, require stricter security compared to comfort and telematics domains \cite{wu2019survey}.
\end{itemize}

\subsubsection{What are the metrics that are based on IVN hardware constraints}
The embedded ECUs and networks in cars differ significantly from traditional IT systems with high-speed Ethernet connections \cite{fallstrand2015applicability}, making it crucial to consider their resource requirements when designing IDS solutions. However, since most ECUs are proprietary with limited documentation on their architecture, analyzing their internal configuration is challenging and may require backdoor reverse engineering. Although this paper will not delve into the intricacies of ECU resource needs due to the unavailability of relevant information, it aims to provide relevant evaluation guidelines from a broad perspective suitable for IVNs with limited resources.

\begin{itemize}[leftmargin=2em]
    \item {\em CPU and memory overheads:} The CPU and memory resources of most ECUs are limited. For example, some low-end ECUs have 8-bit microcontrollers that run at 20MHz with only 32kB memory and 1kB of RAM \cite{sagstetter2013security}. These constraints make it difficult for resource-intensive IDS applications in IVNs to perform real-time functions. As a result, IDS methods that rely on computationally intensive ML/DL techniques may not be suitable for meeting the real-time demands of many of today's cars.
    
    \item {\em Network bandwidth overheads:} To ensure that real-time message delivery is not compromised, it is essential to analyze network bandwidth usage in an IVN since the multiple distributed ECUs use up a significant portion of the network capacity.
    
    \item {\em Power consumption:} The energy consumption of an IDS solution in IVNs is a critical metric to evaluate as automobiles have a limited amount of energy to run.
\end{itemize}

\subsubsection{What are the realistic threats to UAV Networks}
UAVs are vulnerable to attacks as their wireless communication medium is open to exploitation. Attackers can use different techniques to gain control of a UAV network and extract critical mission information. This section concentrates on the most feasible threats to UAVs identified in the literature.

\begin{itemize}[leftmargin=2em]
    \item {\em DoS and DDoS attacks:} A DoS attack involves sending a barrage of requests to the UAV network channel, causing an interruption in normal transmission and resulting in a network outage  \cite{choudhary2018internet}. DDoS attacks, on the other hand, are carried out by multiple actors, who may use distributed entry points to gain access to the network \cite{mairaj2022game}.
    
    \item {\em Jamming attacks:} Jamming is a type of denial-of-service attack that poses a severe danger to UAV networks by impairing network availability \cite{slimane2022light,7414043,antijamming:2016}. In a jamming attack, a malicious user floods the UAV network channels with higher-power radio signals to disrupt ongoing conversations. Control signals or fake GPS signals could occasionally be used as jamming signals \cite{guo2020vulnerabilities}.
    
    \item {\em GPS spoofing attacks:} An attacker can create and transmit fake GPS signals. It can also alter the content of intercepted signals or use a GPS signal generator to produce spoofing signals \cite{choudhary2018internet}.
    
    \item {\em Message forgery and replay attacks:} A forgery attack would aim to disrupt the stability of UAV networks by inserting fabricated information. The fake message may involve sensory data or control signals used to synchronize the operations of the UAV network. Replay attacks, on the other hand, involve replacing legitimate sensory data with previously intercepted data to impair the network's functioning and cause instability \cite{sanchez2019frequency}.
\end{itemize}

%\begin{figure*}[ht]
%\centerline{\includegraphics[width=1.0\linewidth]{figures/uav_architecture.png}}
%\caption{General architecture of an Unmanned Aerial Vehicle as illustrated in \cite{mekdad2021survey}}
%\label{fig3}
%\end{figure*}

\subsubsection{Are current UAV architectures fit to meet the requirements for an IDS}
The internal architecture and characteristics of UAV networks when interconnected are important considerations in the design of practical IDS. This section briefly reviews UAVs' general architecture to aid the readers' understanding of how such architectures dictate the requirements that an IDS must satisfy to be deemed feasible for deployment. 
%A high-level architecture of a UAV is depicted in \autoref{fig3}.

\begin{itemize}[leftmargin=2em]
	    \item {\em Flight controller:} This serves as the CPU of the UAV and acts as an intermediary between the software and the drone's onboard electronics. The memory and CPU capacity of this microcontroller unit may differ depending on the application. However, UAVs are generally resource-limited, although, this is fast changing with rapid technological advancements.
	    \item {\em Ground control station (GCS):} The GCS is a facility on land where human operators provide control signals for the launch, command, and monitoring of a UAV network during operations \cite{yaacoub2020security}. Communication with the UAV swarm must have low-latency and is usually done remotely. An intrusion into the GCS could allow an attacker to take control of the network, putting the UAVs at risk. Therefore, securing the link between the GCS and UAVs is crucial for the network's safety and security.
	    \item {\em Wireless communication module:} The data link module allows for the flow of information between the UAVs and GCS via a transmitter and receiver connected to the flight controller board. The choice of wireless link depends on the operational range of the UAVs, with cellular radio waves for visual line of sight (LOS) and satellite communications for below-visual LOS \cite{yaacoub2020security}.
	    \item {\em Sensors and actuation:} UAV sensors provide location, speed, and height estimates by measuring and ranging surrounding objects. The Flying Controller processes and sends the data to the actuator for actuation during flight \cite{mekdad2021survey}. An attack on the sensor can disrupt the network, especially if the UAV is a relay cluster head.
	    
     \item {\em Power supply:} UAVs are typically powered by lithium polymer batteries. These batteries provide a limited power supply for the UAV system \cite{mekdad2021survey}, depending on the battery's charge and the flight's objective.
\end{itemize}

\subsubsection{What are the metrics that are based on UAV hardware constraints} Based on the hardware architecture described in the previous section, the key considerations for implementing practical IDS for UAV networks are summarized as follows:

\begin{itemize}[leftmargin=2em]
    \item {\em Computational and memory costs:} To ensure that the system security is not compromised, any additional components added to UAVs must have minimal resource demands that fit the hardware capability of the UAV. Since UAVs have limited compute and memory resources, it is essential to strike a balance between detection accuracy and processing/storage overheads associated with IDS.
    \item {\em Network bandwidth costs:} An IDS's efficient utilization of available bandwidth is critical in UAV networks due to limitations caused by the wireless communication channel's capacity, UAV velocity, error-prone wireless connections, and lack of security with broadcast communications, all of which restrict the size of available bandwidth \cite{hentati2020comprehensive}. This limitation is further compounded by the presence of multiple applications that compete for the limited bandwidth available in the network.
    \item {\em Energy costs:} An energy-efficient IDS is key in UAV networks as they are powered by rechargeable batteries with limited energy resources \cite{hentati2020comprehensive}. The energy resources of the individual units determine the UAV network's power when deployed, and inefficient IDS can quickly drain them. Therefore, IDS design should prioritize low power consumption to extend UAV networks' lifetime while maintaining high detection accuracy.
\end{itemize}

\subsubsection{What are the metrics that measure the effectiveness of attack detection in IVNs and UAV Networks} 
The performance of AI/ML-based IDS solutions is typically evaluated using accuracy, recall, precision, and F1-score metrics \cite{zaman2018evaluation}. While detecting attacks with high accuracy is important, it is equally crucial to ensure that this performance can be replicated in practical situations. Therefore, when designing IDSs for IVNs and UAV networks, additional design criteria must be considered. To serve as a reference for evaluating such solutions, the following requirements are developed as a baseline.

\begin{itemize}[leftmargin=2em]
    \item {\em Detection accuracy:} Accuracy is a valuable metric to evaluate the performance of an IDS solution as it shows the proportion of correctly classified instances out of the total instances. When the dataset is balanced, accuracy is sufficient to determine the effectiveness of the IDS. However, for imbalanced datasets, F1-score is considered a better measure of accuracy.
    
    \item {\em False positive and false negative rates:} False Positive rates (FPR) and False Negative rates (FNR) in IVN and UAVs can have catastrophic effects due to their safety-critical nature. Falsely classifying legitimate traffic as an attack can harm passengers or fail the UAV's mission \cite{schell2021assessment}, while missing an actual attack can lead to accidents or mission failures. Therefore, keeping the FPR and FNR as low as possible is important for reliable IDS performance.
    
    \item {\em Response time:} The training time is the time it takes to train a learning-based IDS, while the time it takes to classify traffic is the prediction time. Although accuracy is important, a high detection time can be detrimental, as the attacker may have already caused damage \cite{al2019intrusion}. Moreover, if the training time is too long, the IDS model may not detect new types of attacks. To optimize IDS design, time complexity, and placement strategies must be considered in light of resource constraints on vehicles' ECUs and UAVs.
    
    \item {\em Detection time:} This metric measures the time taken by the IDS to detect an attack. A shorter detection time can help minimize the damage caused by an attack.
    
    \item {\em Mean Time Between False Alarms (MTBF):} This metric measures the average time between false alarms generated by the detection system. A higher MTBF indicates a more accurate detection system.
    
    \item {\em Throughput:} IDSs should not create network bottlenecks \cite{al2019intrusion}. This means that the data throughput of the IVNs and UAV networks must be considered to ensure that there is enough bandwidth to accommodate the IDS overhead. This is critical since some applications are time-sensitive (e.g., vehicular safety applications that require delays of less than 300ms \cite{etsi103}).
    
    \item {\em Model update:} It is not enough to develop a functional IDS; it is imperative to ensure that it can be updated effortlessly to detect new types of attacks without compromising security or causing delays in the update process. Moreover, these updates should be based on feedback from actual network samples.
    
    \item {\em Privacy:} To prevent unauthorized disclosure of private vehicle data to external systems, IDS solutions in IVNs and UAVs should not allow any leakage through their different connection interfaces. This requirement becomes even more critical when the IDS is connected to a cloud backend for incident analysis and update \cite{schell2021assessment}.
\end{itemize}

\subsubsection{What other considerations should influence IDS development for IVN and UAVs} Some additional (secondary) metrics that may be useful to consider when evaluating IDS for IVNs and UAV networks include:

\begin{itemize}[leftmargin=2em]
    \item {\em Price constraints:} The costs of integrating an IDS solution for IVN and UAVs should be a key consideration because a system's economic viability is pivotal to its implementation. A system that requires significant redesigns of current hardware or software, for instance, incurs additional costs and is, therefore, less attractive to manufacturers (or the owners of the IDS solution is implemented as an aftermarket product) \cite{fallstrand2015applicability}.
    
    \item {\em System complexity and ease of deployment:} This is closely related to the last point as increased complexity often leads to higher costs. Desirable systems are those that can easily be deployed, audited, and maintained.
    
    \item {\em Flexibility and scalability of solution:} Flexibility refer to the IDS solution's ability to adjust to various situations that may arise in IVN and UAV networks. It measures how adaptable the solution is to different electronic platforms that have many variations and how well it manages ECU/sensor software upgrades, replacements, and other tasks \cite{stachowski2019assessment}. When designing these systems, scalability is critical, and key factors to consider include \cite{stachowski2019assessment}: (1) Can the solution be easily updated in the future? (2) Is it possible to extend it to detect threats from outside the vehicle? (3) Does it allow for independent upgrades to the IDS software and its configuration?
    
    \item {\em Forensic capabilities:} An effective IDS for IVNs and UAV networks should answer important questions about its forensic capabilities, such as detecting anomalies, effectiveness in detecting zero-day attacks, and available features of cloud-based services \cite{stachowski2019assessment}.
    
    \item {\em Recovery from attacks:} Effective recovery after attacks is often neglected in the IDS solutions proposed in the literature. However, assessing the extent of an intrusion and evaluating its potential impact on the network to recover the IDS effectively from attacks through damage assessment and remediation is equally critical.
    
    \item {\em Relevancy of dataset:} Due to the high costs associated with real-world experiments, IDSs evaluated using synthetic datasets obtained from other network domains or through simulation and testbed evaluations should consider the relevance and generalization of the dataset feature across various real-world scenarios. Models that overfit to simulated environments may not be efficient and adaptable to practical applications.
    
\end{itemize}

\autoref{tab:metrics_ivn} gives a summary of the different feasibility requirements surveyed in this paper. From a high-level perspective, the most critical factors necessary to evaluate the feasibility of the latest IVN and UAV IDS proposals are defined.

\renewcommand{\arraystretch}{1.5}
\begin{table*}[htb]
\caption{\small Summary of metrics useful for assessing IDS and MDS feasibility in networked autonomous systems}
\begin{center}
\begin{tabulary}{\linewidth}{LLLL}
	\hline
	Metrics Category &  & Metric & Rationale \\
	\hline
	\multirow{9}{*}{\parbox{4cm}{Attack Detection}} & 1 & Detection Accuracy & How often the IDS classifies network traffic correctly? \\
	& 2 & False Positives Rate & Misclassification frequency of normal traffic as an attack? \\
	& 3 & False negatives Rate & Misclassification frequency of attack traffic as benign? \\
	& 4 & Training Time & How long does it take to train the IDS/MBD? \\
	& 5 & Detection Time & How long does it take the IDS/MBD to detect an attack? \\
        & 6 & MTBF & How long does it take between consecutive false alarms? \\
	& 7 & Throughput & What is the impact of IDS on vehicle network traffic data rates? \\
	& 8 & Model update & How does the IDS update and how long does it take? \\
	& 9 & Privacy & How does the IDS protect the private data of the vehicle? \\
	& 10 & Relevancy of datasets & How relevant is the evaluation dataset for developing the IDS? \\
	\hline
	\multirow{5}{*}{Hardware Constraints} & 11 & Processing costs & How do the IDS computational needs impact the vehicle's real-time needs? \\
	& 12 & Memory costs & What are the effects of IDS memory usage on a vehicle's real-time requirements? \\
	& 13 & Bandwidth costs & How does the IDS impact the latency requirement of the vehicle? \\
	& 14 & Energy costs & How does the IDS add to the vehicle`s energy demands? \\
	& 15 & Ease of Deployment & What effort and time is required to configure and use the IDS on the vehicle? \\
	\hline
	\multirow{6}{*}{Secondary Factors} & 16 & Price constraint & Does the IDS add to the production costs of the vehicle? \\
	& 17 & System complexity & Does the IDS complexity add any substantial costs to the vehicle? \\
	& 18 & Flexibility & How adaptable is the IDS to different environments? \\
	& 19 & Scalability & Is the IDS future-proof? \\
	& 20 & Forensics capabilities & How does the IDS account for undetected attacks? \\
	& 21 & Recovery & How does the IDS recover after failure? \\
	\hline                                              
%\end{tabular}
\end{tabulary}
\end{center}
\label{tab:metrics_ivn}
\end{table*}

\subsection{Feasibility Requirements for IDS and MDS in V2X Communications} 
This section provides answers to the same questions raised in the preceding section to establish metrics for evaluating the feasibility of IDS/MDS proposals in V2X communications.

\subsubsection{What are the real threats to V2X communication} Due to various external entities that a vehicle needs to interact with, some of the attacks that have already been discussed in this paper for IVNs also impact V2X communications, as well as a few others that are presented next:

\begin{itemize}[leftmargin=2em]
    \item {\em Denial-of-Service (DoS) and Distributed DoS attacks:} DoS attack is mostly aimed at denying access to a system resource. In the context of V2X, an attacker initiating DoS can try to shut down the network between vehicles and roadside infrastructures, denying the vehicle access to critical road safety information. Multiple nodes in V2X can launch a distributed DoS attack to deny vehicular access and make critical network infrastructure unavailable for legitimate access. Various forms of DoS and DDoS attacks have been reported in the literature, including JellyFish, intelligent cheaters, and flooding \cite{sakiz2017survey}.
    
    \item {\em Replay attacks:} In replay attacks, an attacker can resend messages previously transmitted by other vehicles, infrastructure, and pedestrians in an attempt to disrupt traffic flow and cause receiving vehicles to improperly react to non-existing road conditions\cite{lu20195g, lastinec2019analysis}.
    
    \item {\em False information attacks:} A malicious vehicle might generate false information, such as BSMs or sensory data, and broadcast it to the network. This could be done with the aim of causing traffic jams or accidents. This attack can take the form of fake position coordinates, fake speed, false acceleration, or false deceleration messages broadcast to other vehicles nearby \cite{lastinec2019analysis}.
    
    \item {\em Sybil and Blackhole attacks:} Sybil attacks create multiple fake identities to send disruptive messages to other network nodes, while Blackhole attacks manipulate nodes to drop packets instead of forwarding them. Mitigation methods for Sybil attacks include identification and authentication-based approaches \cite{dibaei2019overview, lastinec2019analysis}.
\end{itemize}

\subsubsection{What other considerations should influence IDS/MDS development for V2X communications}

\begin{itemize}[leftmargin=2em]
    \item {\em System resources:} Vehicles have limited processing, memory, and network resources \cite{den2018security}. Thus, IDS/MDS may face resource constraints when filtering out network traffic for attacks or misbehavior due to the strict time requirements of vehicular applications and to the highly dynamic network topology.
    
    \item {\em Real-time constraints:} To design effective IDS solutions for V2X communications, it is important to address strict latency constraints. Outdated/delayed information might not be of use, and hence, complex learning algorithms with high computational demands may not be suitable for real-time applications \cite{el2020cybersecurity}.
    
    \item {\em Threat robustness:} Designing IDS/MDS to identify and detect zero-day attacks is crucial as relying only on known attacks or misbehavior signatures is insufficient. V2X vehicles have various entry points that attackers constantly exploit, making it too difficult to anticipate all possible risks.
    
    \item {\em Flexibility:} IDS/MDS deployed in V2X environments need to be adaptable and reliable, considering the numerous devices and protocols used. Additionally, since V2X networks are dynamic, and network connections exist for a limited time, IDS/MDS should be capable of adjusting to filter traffic, even when deployed in untested scenarios.
  
    \item {\em Privacy:} Another concern for V2X communications is privacy, because vehicles generate sensitive data and send it to other network participants. Identity and location information theft may occur if IDS/MDS techniques cannot ensure that authorized users' data are not exposed when in operation and during model updates.
\end{itemize}

\section{Review of Current IDS/MDS Solutions}
\label{sec:evaluation}
This section reviews and analyzes recently proposed IDS and MDS solutions for VANETs and AV/UAV networks, focusing on the relevant metrics discussed in the preceding section. Any missing metric is also discussed; Tables \ref{tab:ivnrev1}-\ref{tab:uavrev2} are used to summarize our evaluation of the reviewed works.

\renewcommand{\arraystretch}{1.4}
\begin{table*}[htb]
\caption{\small IVN-IDS feasibility metrics assessed in current papers}
\begin{tabulary}{\linewidth}{LCCCCCCCC}
%{p{2.8cm}CCCCCCC}
    \toprule \toprule
    & FPR & FNR & Accuracy & Threat Model & Unseen Attack & Latency(ms) & Placement & Validation \\ 
    \hline
    Zhang et al. \cite{zhang2022hybrid} & $<0.1\%$ & n/a & $>99\%$ & Yes & n/a & 0.55 & C. Gateway & R. Traces \\ \hline
    Hossain et al. \cite{hossain2020effective} & $<0.002\% $& $<0.028$\% &  $>99.9\%$ & Yes & n/a & n/a & n/a & R. Traces \\ \hline
    Lin et al. \cite{lin2020evolutionary} & $<12.89\% $& $<10.92\%$ & $>89\%$ & No & n/a & n/a & n/a & R. Traces \\ \hline
    Hanselmann et al. \cite{hanselmann2020canet} & n/a & n/a & $>99\%$ & Yes & Yes & n/a & n/a & R\&S Traces \\ \hline
    Tariq et al. \cite{tariq2020cantransfer} & n/a & n/a & $>90\%$ & No & Yes & n/a & n/a & R. Traces \\ \hline
    Song et al. \cite{song2020vehicle} & $<0.18\%$ & $<0.35\%$ & $>90\%$ & No & No & 5/6.7 & n/a & R. Traces \\ \hline
    Minawi et al. \cite{minawi2020machine} & $0$ & n/a & $>95.7\%$ & No & n/a & n/a & E-ware & R. Traces \\ \hline
    Zhu et al. \cite{zhu2019mobile} & n/a & n/a & $>90\%$ & Yes & n/a & 0.61 & MEC & R. Traces \\ \hline
    Agrawal et al. \cite{agrawal2022novelads} & n/a & n/a & $>99.9\%$ & Yes & n/a & 128.78 & n/a & R. Traces \\ \hline
    Desta et al. \cite{desta2022rec} & n/a & n/a & $>97\%$ & n/a & n/a & 117 & n/a & Traces \& PoC \\ \hline
    Cheng et al. \cite{cheng2022tcan} & $>0.15\%$ & n/a & $>99\%$ & Yes & No & 3.4 & n/a & Traces \& PoC \\ \hline
    Xie et al. \cite{xie2021threat} & n/a & n/a & $>99\%$ & No & n/a & 0.09 & n/a & R. Traces \\ \hline
    Xun et al. \cite{xun2021vehicleeids} & $<0.1\%$ & $<0.21\%$ & $>97\%$ & Yes & n/a & n/a & G/E-ware & R. Traces \\ \hline
    Javed et al. \cite{javed2021canintelliids} & n/a & n/a & $>93\%$ & No & n/a & n/a & n/a & R. Traces \\
    \bottomrule \bottomrule
\label{tab:ivnrev1}
\end{tabulary}
\begin{minipage}{18cm}
\vspace{-0.3cm}
\begin{center} \small n/a: not applicable; R: Real; R\&S: Real and synthetic; C: Central; G: Gateway; E-ware: External Hardware; PoC: Proof of Concept \end{center}
\end{minipage}
\end{table*}

\renewcommand{\arraystretch}{1.4}
\begin{table*}[htb]
\caption{\small Overlooked IVN-IDS feasibility metrics in current literature}
\begin{tabulary}{\linewidth}{LCCCCCC}
%{p{2.8cm}CCCCCCC}
    \toprule \toprule
    & Privacy Preservation & Comm. Overhead & Compute Overhead & Memory Overhead & Energy Costs & Model Update \\
    \hline
    \cite{zhang2022hybrid}; \cite{hossain2020effective}; \cite{lin2020evolutionary}; \cite{hanselmann2020canet};  \cite{tariq2020cantransfer}; \cite{song2020vehicle}; \cite{minawi2020machine};  \cite{zhu2019mobile};  \cite{agrawal2022novelads}; \cite{desta2022rec};  \cite{cheng2022tcan}; \cite{xie2021threat}; \cite{xun2021vehicleeids}; \cite{javed2021canintelliids} & n/a & n/a & n/a & n/a & n/a & n/a \\
    \bottomrule \bottomrule
\label{tab:ivnrev2}
\end{tabulary}
\begin{minipage}{18cm}
\vspace{-0.3cm}
\begin{center}\small n/a: not applicable \end{center}
\end{minipage}
\end{table*}

\subsection{IVN Intrusion Detection Systems (Tables~\ref{tab:ivnrev1} and~\ref{tab:ivnrev2})}
Zhang et al. \cite{zhang2022hybrid} proposed a hybrid intrusion detection system that combines rule-based and ML-based techniques to achieve efficient and accurate intrusion detection in IVNs. In the first stage, the rule-based technique acts by the rules defined by the authors in \cite{zhang2022hybrid}. The attacks evading this stage are forwarded to the DNN-based classifier in the second stage. The proposed model was trained and evaluated with real data collected from four distinct makes of automobiles. Although the proposed system achieved high detection accuracy on the evaluated dataset, the authors in \cite{zhang2022hybrid} did not clarify the rationale for the rules selection in the first stage. Moreover, important feasibility metrics such as false-negative rate, communication cost, computing overhead, and privacy preservation capabilities were not considered in their evaluation.

Hossain et al. \cite{hossain2020effective} proposed a CNN-based IDS for the CAN bus system and generated an evaluation dataset of CAN messages obtained from three different vehicle models. By evaluating binary and multiclass classification for the three attack types covered in their study, the authors achieved high detection accuracy. However, the proposed dataset is likely too simplistic to capture various practical attack scenarios. Furthermore, rather than presenting metrics needed to determine the feasibility of the proposed method, more emphasis was placed on discussing the methodologies of IDS development.

Lin et al. \cite{lin2020evolutionary} introduced a DL-based denoising autoencoder for extracting and learning features based on packet transmission periodicity for the detection of CAN bus traffic anomalies. The authors collected real CAN traffic from a vehicle to evaluate their solution. Although the proposed IDS was tested using data collected from other vehicles and showed good detection accuracy, only three attacks were modeled.

To capture the temporal dynamics of the CAN bus, the authors in \cite{hanselmann2020canet} introduced CANet, which is based on an LSTM ML algorithm. By using an unsupervised learning approach, the proposed system was designed to detect both known and unknown attacks. Real and synthetic CAN data containing five different attack types were used to evaluate the proposed method.  However, the memory costs were analyzed only analytically (the number of neural network parameters) without experimental insights.

Tariq et al. \cite{tariq2020cantransfer} proposed CANtransfer, a convolutional LSTM-based model that utilizes transfer learning to detect unknown attacks in IVNs. The CAN traffic data from two vehicles are used to capture three types of attacks. Although the detection accuracy appears promising for both known and unknown attacks, the authors did not provide actual numerical value metrics such as processing delays, false alarm rates, and false-negative rates. This makes it difficult to assess the practicality of the proposed system.

Using real traces obtained from injecting normal and attack traffic into the CAN bus, Song et al. \cite{song2020vehicle} proposed and tested the performance of a deep CNN-based IDS. Their findings on five different attack types demonstrated that the proposed model achieves high detection accuracy and low false positive and false negative rates. The authors also estimated that it takes 5ms on a 2.3GHz Intel Xeon CPU and 6.7ms on an Nvidia Tesla K80 GPU to predict a traffic sample using their proposed approach. However, the simplicity of the message injection attacks evaluated in their dataset casts doubt on their proposal's practicality in the real world. Furthermore, the proposed system cannot detect attacks that have never been seen before, and details on how to go about updating the detection model were not provided.

Minawi et al. \cite{minawi2020machine} proposed a ML-based IDS for the CAN bus. The proposed IDS consists of three different inputs, threat detection, and alerting layers. The authors considered and tested different ML algorithms using the car hacking dataset containing real CAN bus messages and various injection attacks. Although the authors in \cite{minawi2020machine} obtained good results in the detection accuracy and false alarm rate, they did not provide details on the false negatives rate, a metric that is as critical as the others presented. Moreover, the proposed IDS placement as external hardware attached to the OBD II port might incur additional costs that the vehicle manufacturers or owners may be unwilling to bear. Similarly, the execution time metric evaluated for the proposed system is not a true measure of the computational performance of the proposed IDS.

A distributed long-short-term memory (LSTM) framework for IVN anomaly detection is proposed by Zhu et al. \cite{zhu2019mobile}. To detect anomalous messages on the CAN bus, the proposed method utilizes multi-dimensional temporal and data properties. They also proposed deploying the proposed IDS on a mobile edge to make use of the additional computational resources available there. However, offloading IDS  operations to the edge could result in considerable delays, making real-time attack detection impossible. Furthermore, given that the proposed IDS only achieved 90\% accuracy on a single test instance, it is hard to predict if it can perform well when applied to other vehicles.

Agrawal et al. \cite{agrawal2022novelads} proposed a DL-based IDS incorporating thresholding and error reconstruction for detecting attack traffic on the IVN. The authors used the car hacking dataset, which contains three attack types, to train and test multiple LSTM architectures and compare their performance. They achieved high attack detection accuracy but mostly at the cost of large computational overheads and a 128ms detection latency. In addition, the authors did not consider other important metrics highlighted in Table \ref{tab:metrics_ivn} to assess the real-world feasibility of the proposed anomaly detection system.

Desta et al. \cite{desta2022rec} introduced a CNN-based IDS that predicts attacks on the CAN bus using temporal relations of CAN messages. The IDS predicts attacks online and offline, and was trained on a dataset with four attack types collected from a real car. The proposed IDS was tested on an Nvidia Jetson TX2 vehicle-class device with an external CAN transceiver connected to the car's CAN bus to demonstrate its proof-of-concept. However, the proposed IDS has some limitations. Firstly, implementing it using an external device to reduce detection latency may result in additional costs that do not meet the economic needs of automakers and buyers. Secondly, it is vehicle-specific and cannot be utilized in other makes. Lastly, updating the proposed IDS online is not possible once it is deployed because the CNN model must be trained from scratch when it significantly diverges from its training dataset.

The TCAN-IDS technique proposed by Cheng et al. \cite{cheng2022tcan} trains a CNN-based ML model with global attention to detect attacks on the CAN bus using the temporal characteristics of CAN messages. On multi-featured temporal data, the authors in  \cite{cheng2022tcan} demonstrated that the proposed TCN model outperforms time-series neural networks, extracting better temporal and spatial features of the CAN messages. When they tested the proposed IDS on a dataset containing attacks of four distinct types on an Intel computer and the Jetson AGX Xavier Nvidia vehicle-class device, the authors obtained a good detection accuracy and detection latency. However, since the CNN model's hyperparameters are set during training, updating it to detect new attacks without retraining is difficult. Furthermore, without considering the impact of other vehicle components, it is hard to assess the efficiency of the inference time.

In their work, Xie et al. \cite{xie2021threat} introduced an enhanced deep learning GAN-based model to detect intrusions on the CAN bus. To mimic the real generated CAN message blocks during the detection phase, the proposed model utilizes complex CAN message blocks in the training samples. The GAN discriminator then utilizes the CAN communication matrix to determine whether each message has been tampered with. The performance evaluation of the proposed IDS involved three stages. Firstly, the GAN model was trained offline on an Intel Xeon CPU with an Nvidia GP102 GPU. Subsequently, the GAN model was tested online on a CAN network prototype consisting of three ECUs. Lastly, the GAN model-based IDS was deployed in a real-world setting utilizing a Xilinx Spartan 6 FPGA connected to the CAN system via a vehicle's OBD II connector. After injecting four attacks into the CAN bus, the proposed IDS was able to identify all of them with high detection accuracy. However, it is important to note that other prediction metrics such as the FPR and FNR were not taken into consideration despite their equal importance.

The technique proposed by Xun et al. \cite{xun2021vehicleeids} is an IDS that monitors message transmission on the CAN bus using vehicle voltage signals. The proposed IDS is based on the fact that ECUs employ somewhat different materials in their hardware, causing them to generate different voltage signals during transmission. These signal features were retrieved using statistical principles and a deep neural network model was implemented to learn and identify anomalous voltage signals on the CAN bus. The authors in \cite{xun2021vehicleeids} validated the proposed IDS on two real vehicles to show that it could be applied to various types of cars. The proposed IDS is designed to be used in the CAN bus automotive gateway and as an independent external device for CAN bus monitoring. However, while the proposed IDS is robust and has a high detection accuracy, it cannot detect attacks originating from the vehicle's ECUs. It can only detect attacks coming from outside sources.

Javed et al. \cite{javed2021canintelliids} introduced CANintelliIDS, an IDS for vehicle intrusion detection on the CAN bus. The proposed IDS uses a combination of CNN and attention-based gated recurrent unit (GRU) to detect single and mixed intrusion attacks on the CAN bus. However, in their evaluation of the proposed IDS's performance using a real dataset consisting of four distinct attacks, they did not include essential evaluation metrics such as FPR, FNR, and detection latency that could further enhance the proposed solution's feasibility in real-world scenarios.

\renewcommand{\arraystretch}{1.4}
\begin{table*}[htb]
\caption{\small V2X-IDS feasibility metrics assessed in current literature}
\begin{tabulary}{\linewidth}{LCCCCCCCC}
%{p{2.8cm}CCCCCCC}
    \toprule \toprule
    & FPR & FNR & Accuracy & Threat Model & Unseen Attack & Latency (ms) & Placement & Validation \\ \hline
    Alladi et al. \cite{alladi2021artificial} & n/a & n/a & $>98\%$ & No & n/a & 14.29 & MEC & S. Traces \\ \hline
    Shu et al. \cite{shu2020collaborative} & n/a & n/a & $>97\%$ & No & n/a & n/a & Decentralized & R. Traces\\ \hline
    Nie et al. \cite{nie2020data} & 0 & n/a & $>97\%$ & No & n/a & n/a & n/a & S. Traces\\ \hline
    Goncalves et al. \cite{gonccalves2021intelligent} & $<0.14\%$ & n/a & $>92\%$ & No & n/a & n/a & n/a & S. Traces \\ \hline
    Kosmanos et al. \cite{kosmanos2019intrusion} & n/a & n/a & $>90\%$ & Yes & n/a & n/a & n/a & S. Traces \\ \hline
    Ghaleb et al. \cite{a2020misbehavior} & $<0.11\%$ & $<0.05\%$ & $>90\%$ & Yes & n/a & n/a & n/a & S. Traces \\ \hline
    Yang et al. \cite{yang2021mth} & $<13.822\%$ & n/a & $>75\%$ & Yes & Yes & 0.574/0.509 & Gateway & R. Traces \\ \hline
    Ashraf et al. \cite{ashraf2020novel} & n/a & n/a & $>97\%$ & No & Yes & n/a & Gateway & R. Traces \\ \hline
    Khan et al. \cite{khan2021enhanced} & n/a & n/a & $>98\%$ & No & Yes & 0.02ms & Gateway & R. Traces \\
    \bottomrule \bottomrule
\label{tab:v2xrev1}
\end{tabulary}
\begin{minipage}{18cm}
\vspace{-0.3cm}
\begin{center}\small n/a: not applicable; R: Real; S: Simulated \end{center}
\end{minipage}
\end{table*}

\renewcommand{\arraystretch}{1.4}
\begin{table*}[htb]
\caption{\small Overlooked V2X-IDS feasibility metrics in current literature}
\begin{tabulary}{\linewidth}{LCCCCCCC}
%{p{2.8cm}CCCCCCC}
    \toprule \toprule
    & Privacy Preservation & Comm. Overhead & Compute Overhead & Memory Overhead & Energy Costs & Model Update \\ 
    \hline
    Shu et al. \cite{shu2020collaborative} & n/a & Considered & Considered & Considered & n/a & Yes\\ \hline
    Goncalves et al. \cite{gonccalves2021intelligent} & n/a & n/a & n/a & $<54497$kb & n/a & n/a \\ \hline
    Yang et al. \cite{yang2021mth} & n/a & n/a & n/a & 16.21mb & n/a & n/a \\ \hline
    Khan et al. \cite{khan2021enhanced} & n/a & n/a & n/a & $3655$kb & n/a & Yes \\ \hline
    \cite{alladi2021artificial}; \cite{nie2020data}; \cite{kosmanos2019intrusion}; \cite{a2020misbehavior}; \cite{ashraf2020novel}  & n/a & n/a & n/a & n/a & n/a & n/a \\
    \bottomrule \bottomrule
\label{tab:v2xrev2}
\end{tabulary}
\begin{minipage}{18cm}
\vspace{-0.3cm}
\begin{center}\small n/a: not applicable\end{center}
\end{minipage}
\end{table*}

\subsection{V2X Intrusion Detection Systems (Tables~\ref{tab:v2xrev1} and~\ref{tab:v2xrev2})}
Alladi et al. \cite{alladi2021artificial} presented an IDS for detecting cyberattacks in the IoV network using a deep neural network model deployed on a MEC server connected to the RSU. The proposed system employs two classification techniques, one of which generates time sequences from network broadcast messages. The second method is based on classifying network traffic using image representations of these time sequences. The authors showed the feasibility of the proposed IDS on a Raspberry Pi 3B by simulating four distinct Deep Learning Engines (DLEs) using the Veremi Extension dataset, which contains traces collected from simulated vehicular network instances. The authors in  \cite{alladi2021artificial} also proposed that the DLEs be trained on resource-rich cloud servers, with the time-critical prediction jobs being performed on locally deployed MEC servers connected to the RSUs. Although they excelled in terms of detection accuracy and prediction time, they did not evaluate the communication overheads associated with the proposed MEC deployment plan.

Shu et al. \cite{shu2020collaborative} proposed a collaborative intrusion detection framework using a multi-discriminator generative adversarial network. This enables multiple distributed SDN controllers to jointly train an intrusion detection model for an entire VANET without directly exchanging sub-network flows. The proposed framework was validated using KDD99 and NSL-KDD experimental datasets and an emulated cloud server with three distributed SDN controllers. While the authors demonstrated the validity of their proposed system in both IID and non-IID contexts with rigorous mathematical proofs, they did not present any numerical results for the metrics evaluated in their experiments.

The authors in Nie et al. \cite{nie2020data} developed a data-driven IDS by examining the link load characteristics of an RSU in an IoV in response to various attacks. They designed and tested different deep neural network models and discovered that a CNN-based model performs best at extracting spatio-temporal features of RSU link loads and thereby identifying intrusions. Although the authors assessed the proposed method's accuracy and compared it to three existing state-of-the-art techniques, they neglected to analyze other critical metrics such as FPR, FNR, and prediction latency that can be used to determine the feasibility of their research.

Goncalves et al. \cite{gonccalves2021intelligent} proposed an Intelligent Hierarchical IDS that splits the network into four levels with several clusters at each level, allowing different ML-based detection approaches to be used. To test the performance of the various ML models evaluated in their work, the authors in \cite{gonccalves2021intelligent} used a simulated dataset reflecting different features and attack types. However, assessing the feasibility of their proposed system is not possible as the authors failed to consider many of the metrics presented in table \ref{tab:metrics_ivn}.

The IDS proposed by Kosmanos et al. \cite{kosmanos2019intrusion} uses a cross-layer set of attributes to train machine learning models to detect spoofing and jamming attacks against connected vehicle platooning communication. The proposed IDS's detection engine is built on Random Forest (RF), k-Nearest Neighbor (KNN), and One-Class Support Vector Machine (OCSVM), as well as cross-layer data fusion methods. The proposed IDS can generate probabilistic outcomes for both known and unknown attacks. The authors used Veins' simulation \cite{sommer2019veins} to evaluate the proposed IDS for a platoon of four vehicles and obtained a dataset with the two attacks studied. Their results showed that the proposed IDS can detect both attacks with precision. However, the proposed IDS is attack-specific and metrics like FPR, FNR, and detection latency were not included.

Ghaleb et al. \cite{a2020misbehavior} proposed a misbehavior-aware on-demand collaborative intrusion detection system based on distributed ensemble learning techniques. Individual vehicles in the proposed system use their data to train a local IDS classifier, which they then share with other vehicles on demand. The best-performing shared IDS are then combined with the locally trained IDS to create an ensemble of classifiers for usage on the local vehicle. While the authors in \cite{a2020misbehavior} used SUMO simulations [104] to simulate various attacks and evaluate the proposed model's basic performance metrics, they did not provide any numerical estimate of the proposed model's computational savings. Additionally, they overlooked the computational and memory overheads.

Yang et al. \cite{yang2021mth} presented a multi-tiered hybrid IDS for the Internet of vehicles that combines signature-based IDS and anomaly-based IDS to detect both existing and zero-day attacks on both intra-vehicle and external vehicular networks. There are four layers of learning models in the proposed multi-tier architecture. The authors utilized two different datasets to assess the performance and efficiency of their proposed IDS. The first dataset was the car hacking dataset, which pertains to the in-vehicle network, while the second dataset was the CICIDS2017 dataset, which contains data on external network traffic. Their evaluation considered different performance metrics and demonstrated the proposed model`s real-world feasibility on a vehicle-level Raspberry Pi 3 IoT device. Although the researchers evaluated the memory requirements of the proposed IDS, metrics such as the FNR and the computational overheads were not considered.

The authors in Ashraf et al. \cite{ashraf2020novel} proposed a DL-based intrusion detection system for identifying suspicious network traffic of V2X communications. Their hybrid IDS utilized the LSTM autoencoder algorithm to identify both new and existing attacks without relying on signatures, and it was designed specifically to detect attacks at the central network gateways of automobiles. To evaluate the IDS, the authors used the vehicle hacking dataset for intra-vehicle networks (IVN) and the UNSW-NB15 dataset for external vehicular networks. However, they only assessed the base learning-based performance metrics in their evaluation, which makes it challenging to fully analyze the proposed IDS's feasibility based solely on the metrics they considered.

Khan et al. \cite{khan2021enhanced} proposed a hybrid IDS that detects internal and external attacks in IoVs. The system uses a bloom filter and a DNN bidirectional LSTM architecture to identify zero-day attacks. The proposed IDS was evaluated on a CAN-based car hacking dataset and the UNSWNB-15 external vehicular network dataset. Results showed that the proposed IDS has a short 7-minute training period, can detect attacks within 0.023 milliseconds, and is a lightweight system with a total memory overhead of 3655kb. However, the study did not include important metrics such as the False Positive Rate (FPR) and False Negative Rate (FNR) for a comprehensive feasibility assessment.

\renewcommand{\arraystretch}{1.4}
\begin{table*}[htb]
\caption{\small V2X-MDS feasibility metrics assessed in current papers}
\begin{tabulary}{\linewidth}{LCCCCCCCC}
%{p{2.8cm}CCCCCCC}
    \toprule \toprule
    & FPR & FNR & Accuracy & Threat Model & Unseen Attack & Latency (ms) & Deployment Technique & Validation \\ \hline
    Hsu et al. \cite{hsu2021deep} & n/a & n/a & $>95\%$ & No & n/a & n/a & OBU & Simulated traces \\ \hline
    Sharma et al. \cite{sharma2020machine} & n/a & n/a & $>85\%$ & No & n/a & n/a  & OBU & Simulated traces \\ \hline
    Wang et al. \cite{wang2021fast} & n/a & n/a & $>90\%$ & No & n/a & $<267$ & n/a & R\&S Traces \\ \hline
    Hawlader et al. \cite{hawlader2021intelligent} & n/a & n/a & $>94\%$ & Yes & n/a & n/a & n/a & R. Traces \& Sim \\ \hline
    Sharma et al. \cite{sharma2021machine} & n/a & n/a & $>98\%$ & No & n/a & n/a & RSUs & R. Traces \& Sim \\ \hline
    Ercan et al. \cite{ercan2021misbehavior} & n/a & n/a & $>84\%$ & No & n/a & Considered & Decentralized & Simulated traces \\ \hline
    Gyawali et al. \cite{gyawali2020machine} & n/a & n/a & $>84\%$ & Yes & n/a & n/a & OBUs & Simulated traces \\ \hline
    Uprety et al. \cite{uprety2021privacy} & n/a & n/a & $>78\%$ & No & n/a & n/a & OBUs & Simulated traces \\ 
    \bottomrule \bottomrule
\label{tab:v2xrev3}
\end{tabulary}
\begin{minipage}{18cm}
\vspace{-0.3cm}
\begin{center}\small n/a: not applicable; R: Real; R\&S: Real \& Simulated; Sim: Simulation\end{center}
\end{minipage}
\end{table*}

\renewcommand{\arraystretch}{1.4}
\begin{table*}[htb]
\caption{\small Overlooked V2X-MDS feasibility metrics in current literature}
\begin{tabulary}{\linewidth}{LCCCCCC}
%{p{2.8cm}CCCCCCC}
    \toprule \toprule
    & Privacy Preservation & Communication Overhead & Compute Overhead & Memory Overhead & Energy Costs & Model Update \\ 
    \hline
    Wang et al. \cite{wang2021fast} & n/a & n/a & n/a & n/a & n/a & Yes \\ \hline
    Uprety et al. \cite{uprety2021privacy} & Yes & Considered & n/a & n/a & n/a & Yes \\ \hline
    \cite{hsu2021deep}; \cite{sharma2020machine}; \cite{hawlader2021intelligent}; \cite{sharma2021machine}; \cite{ercan2021misbehavior}; \cite{gyawali2020machine} & n/a & n/a & n/a & n/a & n/a & n/a \\ 
    \bottomrule \bottomrule
\label{tab:v2xrev4}
\end{tabulary}
\begin{minipage}{18cm}
\vspace{-0.3cm}
\begin{center}\small n/a: not applicable\end{center}
\end{minipage}
\end{table*}

\subsection{V2X Misbehavior Detection Systems (Tables~\ref{tab:v2xrev3} and~\ref{tab:v2xrev4})}
Sharma et al. \cite{sharma2020machine} presented a data-centric misbehavior detection model for the internet of vehicles based on ML. The proposed system integrates plausibility checks with six selected ML algorithms to find the best-performing model for misbehavior detection in VANET. The proposed model was evaluated using the VEREMI dataset with standard machine learning-based metrics. While the authors in \cite{sharma2020machine} proposed deploying the proposed system on the vehicle's OBU to support local detection and privacy preservation, they did not evaluate FPR, FNR, and detection latency metrics, which are necessary to validate that the proposed scheme can meet the real-time and privacy requirements of V2X communications.

The deep learning-based misbehavior detection system proposed by Hsu et al. \cite{hsu2021deep} uses CNN and LSTM-based models to rebuild a vehicle's position information based on incoming BSMs. Based on the received message, an SVM classifier is utilized as a binary classification method to determine if the receiver vehicle has been compromised or not. The proposed MDS was trained and assessed using the popular vehicular reference (VEREMI) dataset, which contains many types of misbehavior in a vehicular setting. The authors evaluated the accuracy of their proposed system but failed to consider other important metrics like FPR, FNR, and prediction latency.

Wang et al. \cite{wang2021fast} introduced an IoV misbehavior detection system based on broad learning and incremental learning methods. Different ML and DL algorithms are trained in the detection module to detect false messages in the proposed system, and the trained model is continuously updated with new data using an incremental learning technique. The authors assessed the performance of their proposed model on three real-world datasets: VeRemi, NGSIM, and PeMS. Specifically, they analyzed the accuracy of their solution and found that it can detect misbehavior in real-time while frequently updating the learning model. However, critical feasibility metrics such as FPR and FNR were not considered.

The authors in Hawlader et al. \cite{hawlader2021intelligent} trained six different ML algorithms in a supervised method to detect position falsification attacks in VANET. The proposed system was trained on the VEREMI dataset, which contains several position falsification attacks, and its performance was validated with Veins simulation. Nevertheless, the authors did not consider FPR, FNR and prediction latency in their study.

The MDS proposed by Sharma et al. \cite{sharma2021machine} combines information from two consecutive BSMs to train multiple ML-based algorithms to detect position falsification attacks in VANETs. The evaluation of the proposed system using the VeReMi dataset showed high detection accuracy in identifying different misbehaviors. The authors also suggested a hierarchical architecture for deployment, where the RSUs are equipped with the detection model, which is different from most other works that proposed deploying the model on OBUs. However, the proposed system's validity was not established using key feasibility metrics such as FPR, FNR, and prediction latency.

In the MDS proposed by Ercan et al. \cite{ercan2021misbehavior}, three new features were integrated for training two separate ML approaches for detecting position falsification attacks in vehicular networks. The performance of the proposed system was examined using the VeReMi dataset, and the results indicate that it outperforms alternative approaches. The authors also proposed a distributed detection method with centralized training. However, they did not evaluate important feasibility metrics including FPR, FNR, and prediction latency.

Gyawali et al. \cite{gyawali2020machine} presented an MDS to detect and prevent false alarms and position falsification attacks in vehicular networks by combining ML and reputation-based methods. The proposed system is trained with a dataset generated via Veins simulations in a realistic vehicular network environment. The proposed system performed well on the public VeReMi dataset according to the standard ML metrics considered. The authors also conducted experiments and examined the proposed system's complexity for real-world applications. Nonetheless, they did not present numerical results to assess the proposed system's computing cost and detection latency.

Uprety et al. \cite{uprety2021privacy} proposed an MDS that addresses the privacy trust issues associated with centralized misbehavior detection techniques. Local ML models are trained using BSM data received on each vehicle and a global federated model is computed using the parameters of the locally trained models. The authors used the public VeReMi dataset to train the local models on the received BSMs as a proof of concept. They used basic ML metrics to assess the performance and obtain numerical results for detection accuracy and communication costs.  However, the VeReMi dataset is not distributed, and due to the system dynamics, the reported results may not be representative of real-world applications.

%\subsection{Inferences Derived from the Review}
\renewcommand{\arraystretch}{1.4}
\begin{table*}[htb]
\caption{\small UAV-IDS Feasibility Metrics Assessed in Current Papers}
\begin{tabulary}{\linewidth}{LCCCCCCCC}
%{p{2.8cm}CCCCCCC}
    \toprule \toprule
    & FPR & FNR & Accuracy & Threat Model & Unseen Attack & Latency (ms) & Deployment Technique & Validation \\ \hline
    Slimane et al. \cite{slimane2022light} & 1.8\% & 2.4\% & $>98\%$ & Yes & n/a & n/a & n/a & Real traces \\ \hline
    Whelan et al. \cite{whelan2022artificial} & n/a & n/a & $>90\%$ & Yes & No & $<124$ms & On-board agent & Real traces \\ \hline
    Moustafa et al. \cite{moustafa2020autonomous} & $<1\%$ & n/a & $~99\%$ & No & No & n/a  & n/a & Real traces \\ \hline
    Abu et al. \cite{abu2022high} & $<1\%$ & $<3\%$ & $>90\%$ & No & Yes & 2.77ms & n/a & Real traces \\ \hline
    Bouhamed et al. \cite{bouhamed2021lightweight} & n/a & n/a & $>85\%$ & No & n/a & n/a & On-board agent & S. Traces \\ 
    \bottomrule \bottomrule
\label{tab:uavrev1}
\end{tabulary}
\begin{minipage}{18cm}
\vspace{-0.3cm}
\begin{center}\small n/a: not applicable; S: Simulated\end{center}
\end{minipage}
\end{table*}

\renewcommand{\arraystretch}{1.4}
\begin{table*}[htb]
\caption{\small Overlooked UAV-IDS Feasibility Metrics in Current Literature}
\begin{tabulary}{\linewidth}{LCCCCCC}
%{p{2.8cm}CCCCCCC}
    \toprule \toprule
    & Privacy Preservation & Comm Overhead & Compute Overhead & Memory Overhead & Energy Costs & Model Update \\ \hline
    Slimane et al. \cite{slimane2022light} & n/a & n/a & Considered & Considered & n/a & n/a \\ \hline
    Bouhamed et al. \cite{bouhamed2021lightweight} & n/a & n/a & n/a & n/a & $22591$mAh & Yes\\ \hline
    \cite{whelan2022artificial}; \cite{moustafa2020autonomous}; \cite{abu2022high}  & n/a & n/a & n/a & n/a & n/a & n/a \\
    \bottomrule \bottomrule
\label{tab:uavrev2}
\end{tabulary}
\begin{minipage}{18cm}
\vspace{-0.3cm}
\begin{center}\small n/a: not applicable\end{center}
\end{minipage}
\end{table*}

\subsection{UAV Intrusion Detection Systems (Tables~\ref{tab:uavrev1} and~\ref{tab:uavrev2})}
A Lightweight Gradient Boosting ML (LightGBM) algorithm to detect subtle jamming attacks on UAV networks was proposed by Slimane et al. \cite{slimane2022light}. The algorithm was trained and evaluated on a dataset consisting of 10,000 samples of both jamming signals and regular traffic. They compared the proposed technique's performance to that of three other ML models and found that the proposed model outperformed the others. The authors considered critical performance metrics previously overlooked by others, but the numerical results for the evaluated metrics were not clearly presented.

The Micro Air Vehicle Intrusion Detection System (MAVIDS) proposed by Whelan et al. \cite{whelan2022artificial} trains a dataset containing normal flight logs to detect GPS spoofing and jamming attacks using unsupervised machine learning techniques. This approach eliminates the difficulty of finding a labeled dataset for training the detection model by using the UAVs' normal flight data. The authors in \cite{whelan2022artificial} employed standard ML metrics and conducted experimental proofs with three microcontroller boards to demonstrate the effectiveness of the proposed solution when deployed aboard a UAV. In addition, they considered prediction latency and throughput as metrics to assess the feasibility of the proposed system, but they also overlooked metrics such as FPR and FNR in their study.

Moustafa et al. \cite{moustafa2020autonomous} presented an autonomous intrusion detection technique for drone networks to detect cyberattacks. To evaluate the effectiveness of their technique, the authors gathered data from a synthetic testbed designed to simulate real-world UAV networks. They then used this data to train and assess the performance of five different machine-learning algorithms. The testbed's dataset included normal traffic as well as three attack events: probing, DoS, and DDoS. Standard ML metrics were evaluated, but little was said about the other metrics needed to evaluate the proposed scheme's computational efficiency and deployment strategy in the real-world.

Abu et al. \cite{abu2022high} developed an autonomous intrusion detection system (UAV-IDS-ConvNet) that uses deep convolutional neural networks to detect attacks on UAVs. The solution proposed by the authors draws on encrypted Wi-Fi traffic data from three distinct types of unmanned aerial vehicles (UAVs). To evaluate their approach, the authors assembled a dataset named UAV-IDS-2020, which features various attacks on UAV networks across both unidirectional and bidirectional communication flows. The dataset encompasses scenarios involving both homogeneous and heterogeneous networked UAVs. However, similar to other prior works, this study did not incorporate metrics to evaluate the computational efficiency of the proposed intrusion detection system.

Bouhamed et al. \cite{bouhamed2021lightweight} proposed a lightweight intrusion detection and prevention system (IDPS) that uses a Deep Q-learning (DQN) model to autonomously detect and prevent network intrusions in UAVs. A periodic offline-learning mechanism for updating the DQN model parameters to learn and adapt to changes in attack patterns was also added to the proposed detection module. A global model is periodically updated with recently exchanged data by the fleet of UAVs in the proposed distributed architecture. The proposed prototype was evaluated on the CICIDS2017 dataset, which included various attack types, and base ML performance metrics, including energy consumption was taken into account. However, other metrics such as FPR, FNR, and detection latency were not considered.

\section{Takeaways From the Reviews}
\label{sec:inference}
In this section, we provide a summary of the key takeaways based on our thorough review of the surveyed papers. 

%\noindent \textbf{Failure to consider false positive and negative rates (FPR and FNR) metrics.}
\subsubsection{Failure to consider and report on false positive rates (FPR) and false negative rates (FNR) metrics}
From Tables  \ref{tab:ivnrev1}-\ref{tab:uavrev1}, it is clear that most of the reviewed works fail to report on the FPR and FNR metrics. Almost all reviewed studies use standard ML metrics only, including accuracy, precision, and recall. But due to the safety-critical implications of an attack misclassification in vehicular networks, reporting the detection accuracy is just as important as  presenting the results of the false alarm and miss rates. Only 31\% (11/36) of the papers reported the FPR and only 17\% (6/36) presented the FNR rates based on our review. Because of these low percentages, many of the current solutions are limited in terms of demonstrating their appropriateness for protecting the devices in the vehicular network. It is also noteworthy that none of the reviewed papers on misbehavior detection in V2X provided results on FPR and FNR (see Table \ref{tab:v2xrev3}).

\subsubsection{Lack of clearly defined benchmark threat models}
The majority of the reviewed papers did not present a clear threat model on which to base their proposed solutions. This is especially evident in studies investigating IDS and MDS solutions for V2X environments, see Tables \ref{tab:v2xrev1} and \ref{tab:v2xrev3}. The reason for this trend is obvious since most V2X-based studies rely on simulated datasets for evaluation. Nonetheless, even for works based on datasets, the necessity of conveying a clear threat model cannot be overstated. Because most publicly available datasets are attack-specific, any system that relies on them must convey a specific-threat model rather than give a misleading sense of generalization. Some of the attacks modeled in the reviewed literature, on the other hand, require further real-world proof to establish their impact and validity. This is particularly beneficial given the lack of real VANET datasets and the fact that extrapolated data from other areas may not be equally applicable to vehicular network settings.

\subsubsection{Lack of after-deployment updating/upgrading mechanisms}
The majority of current solutions do not consider or provide information about their update mechanisms after deployment. This is especially concerning for the IVN IDS papers reviewed (see Table \ref{tab:ivnrev1}) because none of the eleven proposed models can be retrained with new data samples after deployment. Furthermore, researchers have devoted little, if any, attention to mitigating zero-day attacks with the IDS and MDS solutions proposed in recent studies. As our analysis reveals, either no thought is given to designing systems capable of detecting both existing and new attacks, or information on such systems is omitted if they ever exist. That said, it must be acknowledged that the hardware updating problem is not specific to the studied autonomous networked systems but rather exists in other ML and DL domains as well.

\subsubsection{Lack of inference-time and update-time reporting}
Only 11 out of the 36 reviewed articles considered and provided details on the time it takes for their proposed system to classify network traffic. Although some authors assess their model's training and testing duration, such metrics are sometimes not specific and insufficient to accurately determine how such models will respond to real-world scenarios when deployed. While evaluating ML models' training and testing times has its merits, it is far more necessary to provide details on the prediction and model update latency of the proposed systems. Furthermore, it acts as a standard for comparing different solutions for simple adoption by the industry and provides a safety guarantee for the prediction time of an IDS/MDS.

\subsubsection{Absence of placement and deployment strategies}
When it comes to the deployment of the IDS/MDS solutions presented in the literature, it is surprising to see how many implicit assumptions are made. Researchers appear to have given little thought to how these systems will fit into the existing architecture of automobiles, infrastructure, and UAVs. This tendency is particularly alarming for IDS solutions proposed in the literature for IVN (see Table \ref{tab:ivnrev1}), as indicated by our review. Most of the time, researchers do an excellent job of presenting the necessary ideas on the underlying learning models while omitting the finer details of the system's deployment in the real world.

\subsubsection{Absence of energy cost estimation and consideration}
Another noteworthy observation is the absence of an energy consumption metric in virtually all of the reviewed studies. The need for capturing such a metric might be of less importance in IVNs since they are powered by the CAN bus. However, in small (often battery-powered) devices, like drones, it is important to consider power consumption when designing such IDS/MDS solutions. It is rather surprising, though, that no other works, except Bouhamed et al. \cite{bouhamed2021lightweight}, considered and evaluated the amount of power consumed by their solution. This omission is quite striking as learning-based systems are notoriously power-hungry, constantly demanding more power resources from their host device.
     
\subsubsection{Limited consideration of privacy preservation aspects}
In a similar vein, just one paper (Uprety et al. \cite{uprety2021privacy}) among all reviewed papers considered privacy. Most of the other papers do not look into the privacy-preserving features of their proposed solutions. This includes IDS and MDS solutions for V2X networks, which have more interconnections than IVNs and UAV networks, increasing the likelihood of a privacy breach. Because user adoption of these systems is dependent on proof of privacy, researchers need to focus more on incorporating privacy as an evaluation parameter for establishing the feasibility of IDS/MDS solutions in vehicular environments. 

\subsubsection{Failure to consider computational, communication, and memory storage metrics}
Another key finding from our analysis is that most of the reviewed literature gave minimal thought to evaluating the computational, communication, and memory overhead metrics. While these metrics are independent and should be evaluated separately, we have grouped them together because they are interrelated, and evaluating one without the others does not provide an adequate understanding of the proposed system's feasibility. However, according to our study, none of the recent IDS solutions proposed for IVN and UAVs examined any of these metrics, a surprising trend given the resource-constrained nature of these systems and the complexity of developing learning-based detection models. Similarly, for V2X communications, among all the works reviewed, just one approach (Shu et al. \cite{shu2020collaborative}) considered and evaluated all three metrics. This observation calls for more attention from researchers to evaluate these metrics for existing vehicle and UAV capabilities.

\subsubsection{Limited testbed support and proof-of-concept feasibility}
Based on our review, IDS/MDS researchers use essentially two validation approaches. Datasets are either based on simulated data gathered from public sources or on data collected from real devices. The problem with simply depending on datasets to evaluate the effectiveness of the proposed systems has been discussed in the previous sections as well as in the works of \cite{loukas2019taxonomy} and \cite{luo2021deep}. The second approach, which has received the least attention, is the use of testbeds. Only 5 publications (3 V2X MBD and 2 IVN-IDS) performed a proof-of-concept through a test simulation to demonstrate the real-world feasibility of their proposed solution out of the 36 analyzed papers. While the reasons for the low percentage are understandable, greater prototyping of IDS/MDS systems is necessary to increase OEM and user trust.

\subsubsection{Lack of attack prevention and root cause identification}
Among all the reviewed papers covering UAVs, V2X, and IVN, only the work of Bouhamed et al. \cite{bouhamed2021lightweight} tackled the intrusion prevention problem. The majority of current research focuses on successfully detecting anomalies, with little discussion of how to combat the threats detected in the network. Similarly, none of the studies reviewed looked into locating and determining the source of anomalies in impacted devices. A successful IDS should detect abnormal network traffic and be able to pinpoint where it originated and what triggered it. Thus, in vehicular network contexts, research activities on attack prevention and root cause analysis are required for proposing IDS and MDS solutions.

% Advances on the State-of-the Art
%\section{Potential Directions for AL/ML-driven IDS/MDS Research}
\section{AL/ML-driven IDS/MDS Research Directions}
\label{sec:Advances}
This section examines concepts from other areas of ML and DL with the aim of applying them to address the challenges that currently hinder the widespread adoption of learning-based IDS and MDS solutions in real-world scenarios. By reviewing these ideas, we hope to find ways to improve the effectiveness and practicality of these systems.

\subsection{Concept Drift Issues}
As our literature review showed, most proposed solutions are tested on static data, with no regard for the frequent and often dynamic events that happen in the vehicular network. As a result, a system that identifies attacks with a high degree of accuracy may fall short of the required standards soon after deployment. This problem is known as concept drift and has plagued the field of machine learning for a long time, limiting its application for intrusion detection in highly dynamic contexts such as VANETs and UAVs. Concept drift happens when there is a shift in the data distribution upon which a learning-based detection system is modeled. Since real-world data traffic is generated continuously in non-stationary settings, the trained model struggles to respond dynamically to the constantly changing distribution of incoming data streams. That is why a previously tested high-performing model becomes somewhat ineffective after some time. Many existing methods have been proposed to handle this problem, but their performance and prediction accuracy are limited, prompting the development of alternatives. Next, we review and briefly summarize these techniques for interested readers. 

%\paragraph{\textbf{Developing robust learning models with improved drift adaptation performance}}  
~\\ \noindent {\textbf{Developing robust learning models with improved drift adaptation performance.}}  
Leveraging the ideas proposed by Li et al. \cite{yang2021pwpae} on how to design stable and robust drift adaptive models to overcome the performance issues with existing concept drift adaptation methods, an ensemble of base learners can be constructed following the approach outlined next: (i) Collect and sample incoming streams of data to generate a highly representative subset using a clustering algorithm. (ii) Select diverse state-of-the-art drift detection methods and drift adaptation methods to construct an ensemble of high-performing base learners for initial anomaly detection and drift adaptation. (iii) Construct an ensemble model by integrating the prediction probabilities of the base learners based on the methods described in \cite{yang2021pwpae}. (iv) Deploy the final ensemble model for anomaly detection and drift adaptation.

%\subsubsection{\textbf{Developing robust intrusion detection models that are sustainable over time:}}
~\\ \noindent \textbf{Developing robust IDS models that are sustainable over time.}
Andresini et al. \cite{andresini2021insomnia} propose to combine incremental, active, and transfer learning to solve the issues associated with the non-uniformity of data distribution across time. The proposed methodology updates the model using active learning based on only new data samples to maximize the information gain. It also uses the Nearest Centroid Neighbour classifier to reduce the latency caused by manual labeling and updating. Furthermore, it employs permutation-based variable importance measures to explain how drifts present themselves over time. The proposed framework is divided into three phases:

\begin{enumerate}[label=\roman*.]
    \item Labeled traces are used to train the intrusion detection model during the initialization phase, while a separate oracle method is employed to estimate the true labels of the incoming data flow.
    \item In the incremental learning phase, new unlabeled traces are consumed and processed in batches of equal size sequentially. The intrusion detection model and the label estimator are updated continuously with new traces that are unlabeled at inference time and have been class-estimated prior to the model update. 
    \item The explanation phase calculates the global relevance of features to the detection model's decisions in order to track how the model has evolved to meet the network traffic's drifting characteristics.
\end{enumerate}

%% stopped here: Tue
\subsection{Decisions Interpretability}
The lack of interpretation of learning-based detection systems is another deep concern for their application in safety-critical systems. While traditional decision tree-based ML models typically have good interpretability, the reverse is usually true for their DL counterparts, who typically have higher accuracy but are restricted by their theoretical black-box settings. This inability to gain semantic insights behind the predictions erodes users' and automakers' confidence in vehicular network applications. Although there has been considerable research in this field due to the obvious benefits, we next provide a brief overview of some of the core concepts.

~\\ \noindent \textbf{Developing a theoretically provable framework for interpreting IDS decisions.}
Wang et al. \cite{wang2020explainable} presented a framework based on Shapley Additive exPlanations (SHAP) that improve previous methods by providing a solid theoretical foundation applicable to any IDS model. The proposed framework provides local and global interpretability, and local explanations offer details into each feature value specifics that affect the predicted probabilities. The global explanations extract important features from a dataset and investigate the correlations between feature values and certain types of attacks. Interested readers can refer to \cite{wang2020explainable} for details on implementing the proposed framework.

~\\ \noindent \textbf{Producing better explanations and generalizations via domain knowledge.}
Islam et al. \cite{islam2019domain} provided an approach for creating improved explanations and generalizations of model predictions in diverse network intrusion test cases using public domain knowledge based on the principles of Confidentiality, Integrity, and Availability. The infused domain knowledge generalizes the IDS model to detect unknown threats while lowering training time and allowing for better model prediction. A feature generalizer and an evaluator are the two components of the proposed method. The feature generalizer takes the dataset's original features and combines them with domain knowledge to create a compact and understandable feature set. Feature mapping, feature ranking, and feature building are the other tasks covered in this step. On the other hand, the evaluator is responsible for executing and comparing the performance of various types of features.

\subsection{Adversarial Attacks on Detection Models}
Recent research in computer vision has shown that learning-based models are vulnerable to crafted adversarial attacks, resulting in misclassifications and prediction errors. The lack of interpretability of ML/DL models is one clear reason for this. Attackers leverage this knowledge to create adversarial examples to deceive deployed models in autonomous systems. However, developing secure and adversarially-resistant IDS and MDS is crucial due to the obvious safety-critical nature of such autonomous systems.

~\\ \noindent \textbf{Explaining misclassifications using adversarial machine learning.}
Marino et al. \cite{marino2018adversarial} presented a method for generating explanations for a trained classifier's incorrect estimations by determining the smallest changes required to change the model's output. The magnitude of the difference between the modified and original samples is used to visualize the most important features and explain why the samples were misclassified in the first place. This methodology can be adaptable to any classifier with defined gradients and requires no changes to the classifier model. Finally, it gives a mechanism for understanding a classifier's decision boundaries, with the explanation produced from it closely resembling that of a human expert. The stages of the proposed method are briefly outlined below.

\begin{enumerate}[label=\roman*.]
    \item Misclassified samples from the model prediction are collected and fed into the next step.
    \item The samples are modified until they are successfully classified by solving an optimization problem imposed upon them with an adversarial constraint.
    \item The misclassified and corrected samples are visualized using a dimensionality reduction technique.
    \item Explanations are generated by visualizing the difference between the misclassified and modified samples.
\end{enumerate}

\subsection{Realtime Testbed Evaluations}
One of the key issues researchers face when developing IDS and MDS approaches is the lack of real datasets, which is mainly due to the difficulty of launching large-scale tests and data collections. As a result, many of the solutions proposed in the literature are validated using simulated data, which does not accurately represent real-world testing conditions. Given this difficulty, testbed evaluation has been proposed to reproduce the onboard network and components of an external world, representing needed use cases and driving scenarios in controlled settings.

~\\ \noindent \textbf{Developing in-vehicle architecture suitable for real environment trials.}
Jadidbonab et al. \cite{jadidbonab2021realtime} proposed a multi-component testbed representing a flexible and functional in-vehicle architecture for real-world trials of IDS solutions in training, testing, validation, and demonstration. The majority of the vehicle's architecture and CAN-bus network are simulated using the Vector CANoe network simulator. A car simulator, an onboard network simulator, a physical network, a real car's instrument cluster, and spoofing hardware are all part of the proposed testbed. The driving scenarios are generated using the CARLA simulator data to recreate realistic CAN traffic as input into the CAN Bus. The ECUs used in the testbed were virtualized to resemble those found in a real car. According to the testing results, the proposed testbench can discover issues in attack detection that would otherwise go undetected in an offline test environment.

\section{Conclusion}
\label{sec:conclusion}
This survey investigated the feasibility of utilizing AI/ML-based intrusion and misbehavior detection systems in vehicular communication systems for both on-land vehicles and UAVs. Despite extensive research in this field, the authors found that a few solutions have actually been implemented in practical applications, partly due to a lack of emphasis on demonstrating practicality in real-world scenarios. Additionally, the evaluation metrics used in the literature were found to be inadequate in measuring the feasibility of IDS and MDS solutions.

To address these issues, the paper first analyzed relevant factors such as the current architecture and realistic threats in autonomous vehicles, UAV networks, and V2X communication systems. Baseline requirements for evaluating the feasibility of AI/ML-based IDS and MDS approaches were then defined based on this analysis. Recent papers published in top journals over the last three years were reviewed using these metrics, with results revealing that most papers did not sufficiently consider these feasibility metrics.

Based on these findings, this paper proposes improvements to current AI/ML-based IDS and MDS solutions using techniques explored in other machine learning domains. By highlighting the limitations of current evaluation methods and providing ideas for improvement, it is hoped that researchers will focus on making more realistic evaluations of their techniques, ultimately leading to greater adoption by the industry.

\section*{Acknowledgments}
The authors would like to thank Dr. Samson Damilola Fabiyi, Dr. Emmanuel Gbenga Dada, and Mr. Dismas Ezechukwu for their invaluable feedback on this work.

\bibliographystyle{IEEEtran}
\bibliography{References}

\end{document}